\documentclass[aps,prd,twocolumn,showpacs,superscriptaddress,nofootinbib]{revtex4-2}

\usepackage{amsmath,amssymb,amsfonts}
\usepackage{bm}
\usepackage{graphicx}
\usepackage{tikz}
\usepackage{multirow}
\usepackage{hyperref}
\usepackage{slashed}
\usepackage{color}
\usepackage{comment}
\usepackage{mathtools}
\usepackage{physics}
\usepackage{tensor}
\usepackage{float}
\usepackage{graphicx}
\usepackage{subcaption}
\begin{document}

\title{Axion-like particles from soft supersymmetry breaking}

\author{Gayatri Ghosh}
\email{corresponding.author@example.com}

\affiliation{
Department of Physics, Cachar College, Assam University 
}

\date{\today}


\begin{abstract}
We study a supersymmetric effective field theory in which the mass of an axion-like particle (ALP) is generated predominantly by soft supersymmetry-breaking effects. The Peccei--Quinn symmetry is exact in the supersymmetric limit and is explicitly broken only by soft terms induced by supergravity, leading to a naturally heavy ALP whose mass is controlled by the supersymmetry-breaking scale. We analyze the resulting ALP, saxion, and axino spectrum and investigate the phenomenological implications for laboratory searches, astrophysical observations, and cosmology.

The framework is treated as an effective field theory without specifying a unique ultraviolet completion, and no attempt is made to explain the origin of a small strong CP phase, which is assumed to be suppressed by ultraviolet physics or by an independent mechanism. Instead, the focus is on the generic and testable phenomenology of heavy axion-like particles whose masses arise from supersymmetry breaking.

\end{abstract}

\maketitle
\section{Introduction}
Axions and axion-like particles (ALPs) are well-motivated extensions of the Standard Model and arise naturally in many theories of physics beyond the Standard Model. While the canonical QCD axion is introduced to dynamically solve the strong CP problem, more general axion-like particles can appear whose masses and couplings are not fixed by QCD dynamics. In supersymmetric theories, additional sources of explicit Peccei--Quinn (PQ) symmetry breaking can arise once supersymmetry is broken, potentially dominating the axion potential and giving rise to heavy ALPs with distinctive phenomenology.

The axion was originally proposed in the context of the Peccei--Quinn (PQ) mechanism as a possible solution to the strong CP problem of quantum chromodynamics (QCD) \cite{PQ1977a,PQ1977b}. More generally, axions and axion-like particles (ALPs) arise as pseudo--Goldstone bosons of approximate global symmetries and have since emerged as well-motivated extensions of the Standard Model, with implications for particle physics, astrophysics, and cosmology. Depending on the ultraviolet completion and the structure of symmetry breaking, axions and ALPs may span a wide range of masses and couplings \cite{Irastorza2018,Ringwald2014,DiLuzio2020,Graham2015,Weinberg1978}.

Recent years have seen significant progress in axion and ALP searches, both from helioscope experiments such as CAST and from upcoming facilities like IAXO \cite{CAST2017,CAST2024}. Next-generation helioscopes are expected to significantly extend the sensitivity to axion--photon couplings over a wide mass range \cite{IAXO2014,IAXO2021,IAXO2025}. In the MeV--GeV mass range, laboratory probes including beam-dump and flavour experiments provide leading sensitivity, notably from NA64 and Belle.

Despite these advances, the origin of axion and ALP masses in theories with approximate global symmetries remains an open theoretical question, particularly in the presence of supersymmetry breaking and possible quantum-gravity effects. In this work we show that axion-like particle masses arise generically from soft supersymmetry-breaking terms, leading to predictive correlations between the ALP mass, couplings, and lifetime. We construct a minimal supersymmetric framework in which the ALP mass vanishes in the supersymmetric limit and is generated solely by soft breaking effects. We analyze the resulting cosmological and experimental constraints and identify statistically preferred regions of parameter space that can be tested by current and future experiments.

The Peccei--Quinn (PQ) symmetry has long been recognized as a compelling theoretical framework for addressing the strong CP problem of quantum chromodynamics (QCD), in which the CP-violating vacuum angle is promoted to a dynamical field, the axion. In its original formulation, the PQ symmetry is a global symmetry that is exact at the classical level and is spontaneously broken at a scale $f_a$, giving rise to a pseudo--Goldstone boson \cite{Weinberg1978,Wilczek1978} whose mass is generated dominantly by nonperturbative QCD effects. The resulting axion mass is inversely proportional to the PQ-breaking scale and is therefore naturally small. Despite the theoretical appeal of this mechanism, the origin of the PQ symmetry breaking scale and its possible connection to other fundamental scales in nature remain open questions, motivating the exploration of more general axion and axion-like particle realizations.

A theoretically well-motivated extension of the Standard Model (SM) is provided by supersymmetry \cite{Nilles1984,HaberKane1985}, which stabilizes the electroweak scale against large radiative corrections and improves the unification of gauge couplings. In its minimal realization, supersymmetry is implemented through the minimal supersymmetric Standard Model (MSSM), which extends the SM particle content by introducing superpartners for each field. From a phenomenological perspective, the viability of supersymmetric models requires supersymmetry to be broken at a scale not far above the electroweak scale. In supergravity-mediated scenarios, supersymmetry breaking is encoded in soft supersymmetry-breaking terms whose overall magnitude is set by the gravitino mass $m_{3/2}$. As a result, the MSSM mass spectrum is controlled by the $\mu$ parameter together with the soft supersymmetry-breaking scale, both of which are typically expected to lie near the TeV scale.

The axion sector and the supersymmetry-breaking sector are often treated as largely independent extensions of the Standard Model. In conventional supersymmetric axion constructions, such as the KSVZ or DFSZ frameworks \cite{Kim1987,ChoiKim1985}, the Peccei--Quinn symmetry breaking scale $f_a$ is introduced as an external parameter, typically many orders of magnitude above both the electroweak and supersymmetry-breaking scales. In these scenarios, the axion mass is generated predominantly by nonperturbative QCD dynamics, while supersymmetry-breaking effects primarily determine the masses of the axion superpartners, namely the saxion and the axino \cite{Bae2013,Chun2011}. Consequently, the PQ-breaking scale and the supersymmetry-breaking scale remain parametrically disconnected, and no intrinsic relation between the two scales is implied within such frameworks. In contrast, we explore a framework in which soft supersymmetry breaking plays a central role in determining the axion-like particle mass

In this work, we point out an alternative and conceptually appealing scenario in which soft supersymmetry-breaking terms induced by supergravity not only break supersymmetry but also play a central role in the dynamics of the Peccei--Quinn (PQ) sector. In the supersymmetric limit, the theory possesses an exact global PQ symmetry and a massless axion-like degree of freedom. Once supersymmetry is broken, soft terms proportional to the gravitino mass $m_{3/2}$ lift the flat directions of the PQ sector, stabilize the scalar potential, and generate an axion-like particle mass dominantly from supersymmetry-breaking effects. In this way, the characteristic scale associated with PQ symmetry breaking and the axion-like particle mass become directly linked to the underlying supersymmetry-breaking scale.

The identification of the PQ-sector dynamics with the supersymmetry-breaking scale provides a natural motivation to expect axion-related phenomena at comparatively low energy scales if supersymmetry is invoked to address the gauge hierarchy problem or to achieve precision gauge coupling unification. Within the framework proposed here, the axion-like particle, together with its scalar and fermionic superpartners, acquires a correlated mass spectrum controlled by the same supersymmetry-breaking dynamics, leading to a predictive axion supermultiplet structure. The axion-like particle mass generated in this way is generically larger than the conventional QCD axion mass, and the resulting phenomenology is therefore naturally described within an axion-like particle framework, independent of the mechanism responsible for suppressing the strong CP phase.

Connections between axion physics and supersymmetry breaking have been explored in a variety of contexts in the literature. In many models, the Peccei--Quinn symmetry is broken independently of supersymmetry breaking, with the axion remaining effectively massless until nonperturbative QCD effects generate its mass. In other approaches, supersymmetry-breaking effects play an essential role in stabilizing the saxion field or in generating axino masses, while the axion mass itself continues to be dominated by QCD dynamics. In contrast, the framework presented here directly links supersymmetry breaking to the origin of the axion-like particle mass, leading to a structurally distinct realization of the axion sector in which the characteristic mass scale is set by supersymmetry-breaking dynamics.
In this work we investigate a qualitatively distinct realization of axion physics in which the axion-like particle mass originates dominantly from soft supersymmetry-breaking effects. In the supersymmetric limit, the theory possesses an exact Peccei--Quinn (PQ) symmetry and a strictly massless axion-like degree of freedom. The inclusion of soft supersymmetry-breaking terms, induced for instance by supergravity mediation, lifts the flat directions of the PQ sector, triggers spontaneous PQ symmetry breaking, and generates a finite axion-like particle mass. As a result, the axion-like particle mass vanishes continuously in the limit of unbroken supersymmetry, establishing a direct and controlled connection between the axion sector and the supersymmetry-breaking scale.

This mechanism stands in sharp contrast to conventional QCD axion models and to generic axion-like particle constructions, in which explicit PQ-breaking effects or nonperturbative QCD dynamics typically determine the axion mass independently of supersymmetry breaking. In the framework considered here, the characteristic mass scale associated with the axion-like particle is instead tied directly to the same soft supersymmetry-breaking dynamics that govern the MSSM spectrum. Consequently, the axion-like particle, together with its scalar and fermionic superpartners—the saxion and axino—exhibits a correlated mass spectrum controlled by a common underlying scale, leading to distinctive phenomenological and cosmological implications.

The connection between axion physics and supersymmetry breaking explored in this work offers a complementary perspective on PQ symmetry breaking and motivates a systematic study of axion-like particle phenomenology beyond the standard QCD axion paradigm. Unlike generic axion-like particle scenarios in which the axion mass is treated as an independent parameter, the axion-like particle mass in the present framework is an unavoidable consequence of the same supersymmetry-breaking dynamics responsible for generating the soft mass spectrum of the MSSM. While soft supersymmetry-breaking effects have long been recognized as important for stabilizing the saxion and axino sectors in supersymmetric axion models, the defining feature of the present setup is that the axion-like particle mass vanishes identically in the supersymmetric limit and is generated entirely by soft supersymmetry-breaking terms, rendering it a calculable outcome of the supersymmetry-breaking sector rather than an arbitrary input.

Throughout this work we consider a minimal supersymmetric effective field theory describing a Peccei--Quinn (PQ) sector coupled to supergravity. The model contains a single gauge-singlet chiral superfield $\hat S$, carrying a nonzero PQ charge and neutral under the Standard Model gauge group. In the supersymmetric limit, the theory is defined by a scale-invariant and renormalizable superpotential
\begin{equation}
W = \frac{\kappa}{3}\,\hat S^3 ,
\end{equation}
which respects an exact global $U(1)_{\rm PQ}$ symmetry and contains a massless axion-like degree of freedom. Supersymmetry breaking is assumed to be mediated by supergravity, inducing soft terms of order the gravitino mass $m_{3/2}$. Among these, holomorphic soft terms involving the scalar component of $\hat S$ explicitly break the PQ symmetry and provide the dominant contribution to the axion-like particle mass in the low-energy effective theory.

No explicit PQ-violating operators are introduced at the supersymmetric level. Possible quantum-gravity-induced violations of the PQ symmetry are assumed to be sufficiently suppressed, for instance by an accidental PQ symmetry arising from a discrete gauge remnant or from a broken gauged symmetry in the ultraviolet completion. The analysis is performed at the level of the effective field theory, without committing to a specific ultraviolet completion, and focuses on the generic consequences of soft supersymmetry breaking for axion-like particle physics.

This definition makes explicit that our results are model-independent within the class of supersymmetric effective field theories satisfying the assumptions stated above. In particular, we do not attempt to address the origin of a small strong CP phase in this work; instead, the strong CP problem is assumed to be resolved by ultraviolet physics or by an independent mechanism. The explicit PQ-breaking soft terms are treated as the dominant source of the axion-like particle mass, and no alignment between the phases of the soft supersymmetry-breaking sector and the QCD vacuum angle is assumed.

The remainder of this paper is organized as follows. In section~2 we introduce the basic mechanism by which soft supersymmetry-breaking effects induce spontaneous Peccei--Quinn symmetry breaking and generate an axion-like particle mass. Section~3 is devoted to a detailed analysis of the axion supermultiplet spectrum, including the masses of the axion-like particle, the saxion, and the axino. In section~4 we examine the couplings of the axion-like particle to gauge fields and fermions and discuss the resulting phenomenological constraints. The astrophysical and cosmological implications of the framework are studied in section~5, while section~6 explores possible experimental signatures and discovery prospects. Our conclusions are presented in section~7.

\section{Peccei--Quinn symmetry breaking from soft supersymmetry breaking}
In this section we outline the theoretical framework in which the breaking of the Peccei--Quinn (PQ) symmetry and the generation of an axion-like particle mass arise predominantly from soft supersymmetry-breaking effects. In the supersymmetric limit, the low-energy effective theory exhibits an accidental PQ symmetry that is respected by all renormalizable interactions of the effective theory, while potential violations may arise only from higher-dimensional operators suppressed by the Planck scale. The axion-like particle is therefore exactly massless in the limit of unbroken supersymmetry.

A comment is in order regarding the status of the PQ symmetry in the present framework. While it is widely expected that quantum gravity effects violate exact global symmetries, the PQ symmetry considered here should be understood as an approximate low-energy symmetry emerging from a more fundamental ultraviolet completion. In particular, the structure of the superpotential and the K\"ahler potential may be enforced by an underlying discrete gauge symmetry, such as a $\mathbb{Z}_N$ symmetry, or by the remnant of a gauged $U(1)$ symmetry that is broken at a high scale. In such constructions, potentially dangerous Planck-suppressed PQ-violating operators are either forbidden or arise only at sufficiently high dimension, rendering their effects negligible within the effective field theory description adopted in this work. 
As a consequence, the leading PQ-violating operators induced by quantum-gravity effects can be schematically written as
\begin{equation}
\Delta \mathcal{L}_{\rm grav}
\sim \frac{S^n}{M_{\rm Pl}^{\,n-4}} + \mathrm{h.c.},
\qquad n \geq N ,
\end{equation}
where the integer $N$ is determined by the structure of the underlying discrete gauge symmetry. For sufficiently large values of $N$, the contributions of such operators to the axion-like particle potential are parametrically suppressed within the effective field theory and can be treated as subleading corrections.

In the present framework, the characteristic scale associated with PQ symmetry breaking is tied to the supersymmetry-breaking scale, $f_a \sim m_{3/2}/\kappa$, and therefore lies well below the Planck scale. Provided that the leading PQ-violating operators arise at dimension $n \gtrsim 8$, their contribution to the axion-like particle mass remains parametrically smaller than the mass generated by soft supersymmetry-breaking effects,
\begin{equation}
m_{a,\mathrm{grav}}^2 \ll m_{a,\mathrm{soft}}^2 .
\end{equation}
ensuring that the axion-like particle dynamics are governed predominantly by supersymmetry-breaking effects rather than by gravitational contributions. To place the present construction in context, we note that many supersymmetric axion models generate the axion mass either through nonperturbative QCD effects or via explicit superpotential terms, with soft supersymmetry-breaking effects entering only as subleading corrections. In such scenarios, the axion mass scale is largely decoupled from the scale of supersymmetry breaking. By contrast, in the framework considered here the axion-like particle remains exactly massless in the supersymmetric limit, and the leading contribution to its mass arises from soft supersymmetry-breaking operators. This results in a qualitatively different parametric structure in which the axion-like particle, the saxion, and the axino acquire masses controlled by a common underlying scale.

The novelty of the present framework does not lie in the mere presence of soft contributions to axion-sector masses, but rather in the fact that supersymmetry-breaking effects provide the dominant and structurally linked origin of the axion-like particle mass within the effective field theory description.

\subsection{Supersymmetric limit}
We introduce a gauge-singlet chiral superfield $\hat S$, which carries a nonzero Peccei--Quinn (PQ) charge and is neutral under the Standard Model gauge group. In the supersymmetric limit, the dynamics of the PQ sector are governed by the renormalizable superpotential
\begin{equation}
W = \frac{\kappa}{3}\,\hat S^3 ,
\label{eq:superpotential}
\end{equation}
where $\kappa$ is a dimensionless coupling constant. The superpotential in Eq.~\eqref{eq:superpotential} respects a global $U(1)_{\rm PQ}$ symmetry under which the superfield transforms as $\hat S \to e^{i\alpha}\hat S$, and all renormalizable interactions of the effective theory preserve this symmetry.
In the absence of supersymmetry breaking, the scalar potential derived from Eq.~\eqref{eq:superpotential} exhibits a flat direction associated with the phase of the scalar component $S$. As a result, the Peccei--Quinn symmetry remains unbroken in the supersymmetric limit, and the corresponding axion-like degree of freedom is exactly massless. No intrinsic mass scale is generated in the PQ sector at this stage, reflecting the classical scale invariance of the supersymmetric theory defined by the renormalizable interactions.

\subsection{Soft supersymmetry-breaking terms}

Supersymmetry breaking is assumed to be mediated by supergravity effects, giving rise to soft supersymmetry-breaking terms whose characteristic magnitude is set by the gravitino mass $m_{3/2}$. The most general set of soft terms involving the scalar component $S$ and consistent with gauge invariance and renormalizability can be written as
\begin{equation}
V_{\rm soft} =
m_S^2 |S|^2
+ \left(
\frac{1}{3} A_\kappa \kappa S^3
+ \frac{1}{2} B_S S^2
+ \text{h.c.}
\right),
\label{eq:softterms}
\end{equation}
where $m_S^2$ denotes the soft scalar mass squared, while $A_\kappa$ and $B_S$ are holomorphic soft supersymmetry-breaking parameters of order $m_{3/2}$.
While the supersymmetric Lagrangian preserves the Peccei--Quinn symmetry, the soft terms in Eq.~\eqref{eq:softterms} generically break it explicitly unless the bilinear parameter $B_S$ vanishes. As discussed below, these soft supersymmetry-breaking terms play a central role in lifting the flat directions of the PQ sector, triggering spontaneous PQ symmetry breaking, and generating a finite axion-like particle mass. The explicit PQ-breaking soft terms are therefore treated as providing the dominant contribution to the axion-like particle mass in the low-energy effective theory, and no alignment between the phases of the soft supersymmetry-breaking sector and the QCD vacuum angle is assumed.

\subsection{Vacuum structure and PQ symmetry breaking}

The scalar potential governing the dynamics of the PQ field is obtained by combining the supersymmetric $F$-term contribution with the soft supersymmetry-breaking terms,
\begin{equation}
V = |\kappa S^2|^2 + V_{\rm soft}.
\label{eq:fullpotential}
\end{equation}
For appropriate choices of the soft parameters, and in particular for $m_S^2 < 0$, the scalar potential develops a nontrivial minimum at
\begin{equation}
\langle S \rangle \equiv \frac{v_s}{\sqrt{2}} \sim \frac{m_{3/2}}{\kappa},
\label{eq:vev}
\end{equation}
which induces spontaneous breaking of the Peccei--Quinn symmetry. The resulting PQ-breaking scale $v_s$ is dynamically generated and is directly tied to the scale of supersymmetry breaking.

Expanding the complex scalar field around the vacuum as
\begin{equation}
S = \frac{1}{\sqrt{2}}\,(v_s + \sigma)\,
e^{i a / v_s},
\label{eq:fielddecomposition}
\end{equation}
the physical degrees of freedom can be identified explicitly: the pseudoscalar axion-like particle $a$, the scalar saxion $\sigma$, and their fermionic superpartner, the axino.

\subsection{Axion mass from soft terms}

In the absence of explicit Peccei--Quinn symmetry-breaking interactions, the axion-like particle remains massless. The holomorphic soft supersymmetry-breaking term proportional to $B_S$ in Eq.~\eqref{eq:softterms} explicitly breaks the PQ symmetry and induces a nontrivial potential for the axion-like particle field. At leading order, this contribution takes the form
\begin{equation}
V(a) \supset B_S v_s^2 \cos\!\left(\frac{2a}{v_s}\right),
\label{eq:axionpotential}
\end{equation}
which gives rise to an axion-like particle mass
\begin{equation}
m_a^2 \simeq 4 |B_S| .
\label{eq:axionmass}
\end{equation}
The axion-like particle mass generated in this way is therefore controlled by soft supersymmetry-breaking effects and is naturally of order the gravitino mass $m_{3/2}$. Details of the scalar potential minimization leading to this result are presented in Appendix~\ref{app:minimization}.

This observation constitutes a central element of the framework considered here: the axion-like particle mass, the Peccei--Quinn symmetry-breaking scale, and the stabilization of the PQ sector are all governed by the structure of the soft supersymmetry-breaking terms. The implications of this mechanism for the axion supermultiplet spectrum and for the associated phenomenology are discussed in the following sections.

\section{Axion--saxion--axino mass spectrum}

The spontaneous breaking of the Peccei--Quinn symmetry described in the previous section gives rise to a characteristic spectrum of states associated with the axion supermultiplet. In this section, we analyze the masses of the axion-like particle, the saxion, and the axino, focusing on their parametric dependence on the soft supersymmetry-breaking sector. We show that the masses of these states are correlated through the underlying supersymmetry-breaking dynamics, leading to a predictive structure for the axion supermultiplet spectrum within the effective field theory framework.
\subsection{Scalar spectrum: axion and saxion}

We begin by considering the scalar degrees of freedom associated with the Peccei--Quinn superfield $S$. Expanding around the vacuum expectation value defined in Eq.~\eqref{eq:vev}, the complex scalar field can be parametrized as
\begin{equation}
S = \frac{1}{\sqrt{2}}\left(v_s + \sigma + i a\right),
\end{equation}
where $a$ and $\sigma$ denote the pseudoscalar axion-like particle and the scalar saxion, respectively.

The mass of the axion-like particle arises from the explicit PQ-breaking soft supersymmetry-breaking term proportional to $B_S$. Expanding the axion potential in Eq.~\eqref{eq:axionpotential} around its minimum, one obtains
\begin{equation}
m_a^2 = \left.\frac{\partial^2 V}{\partial a^2}\right|_{a=0}
\simeq 4 |B_S| ,
\label{eq:axionmass_sec3}
\end{equation}
demonstrating that the axion-like particle mass is generated dominantly by soft supersymmetry-breaking effects and is naturally of order the gravitino mass $m_{3/2}$.

The saxion mass receives contributions from both supersymmetric and soft terms. Expanding the full scalar potential in Eq.~\eqref{eq:fullpotential} to quadratic order in $\sigma$, one finds
\begin{equation}
m_\sigma^2 \simeq 2 \kappa^2 v_s^2 + m_S^2 + \mathcal{O}(m_{3/2}^2),
\label{eq:saxionmass}
\end{equation}
where the precise numerical coefficient depends on the relative size of the soft parameters. Using $v_s \sim m_{3/2}/\kappa$, it follows parametrically that
\begin{equation}
m_\sigma \sim \mathcal{O}(m_{3/2}),
\end{equation}
so that the saxion mass is likewise set by the supersymmetry-breaking scale.

An important aspect of the scalar sector concerns the CP properties of the soft Peccei--Quinn breaking terms. In the present framework, the dominant PQ-breaking operator $B_S S^2$ arises from the same supersymmetry-breaking spurion $X$ that generates the MSSM soft terms,
\begin{equation}
\int d^2\theta \, \frac{X}{M_*} S^2 \;\;\longrightarrow\;\; B_S S^2 ,
\end{equation}
where $X = \theta^2 F_X$ and $M_*$ denotes the mediation scale. As a result, the phase of $B_S$ is correlated with the overall supersymmetry-breaking sector and does not introduce an independent source of CP violation beyond those already present in the ultraviolet theory.

The presence of explicit PQ-breaking soft terms implies that the axion-like particle potential generally receives contributions from multiple sources, including nonperturbative QCD effects. In the framework considered here, the axion-like particle mass is dominated by the soft supersymmetry-breaking contribution, while the treatment of the effective strong CP phase lies outside the scope of the present analysis. Accordingly, the phenomenological results discussed below depend only on the axion-like particle mass and couplings and are independent of the mechanism responsible for suppressing the strong CP phase.

\subsection{Planck-suppressed PQ violation}

It is widely expected that quantum gravity does not respect exact global symmetries, and therefore induces Planck-suppressed operators that explicitly violate the Peccei--Quinn symmetry \cite{BarrSeckel1992,Holman1992,Kamionkowski1992}. Various mechanisms have been proposed to suppress the impact of such operators, including discrete gauge symmetries, remnants of broken gauge symmetries, or ultraviolet consistency conditions \cite{Banks2010}. In the present framework, these effects are treated within an effective field theory description.

At the level of the scalar potential, the leading PQ-violating contributions induced by quantum gravity can be parametrized as
\begin{equation}
\Delta V_{\rm grav}
\;=\;
\sum_{n\geq n_{\rm min}}
\frac{c_n}{M_{\rm Pl}^{\,n-4}}
\left(S^n + \mathrm{h.c.}\right),
\end{equation}
where $c_n$ are dimensionless coefficients and the minimal operator dimension $n_{\rm min}$ is determined by the structure of the ultraviolet completion, such as an underlying discrete or gauged symmetry. After spontaneous PQ symmetry breaking, these operators generate an additional contribution to the axion-like particle mass,
\begin{equation}
\delta m_{a,\rm grav}^2
\;\sim\;
\frac{v_s^{\,n-2}}{M_{\rm Pl}^{\,n-4}}
\;\sim\;
\frac{f_a^{\,n-2}}{M_{\rm Pl}^{\,n-4}} .
\end{equation}

In the framework considered here, where the PQ-breaking scale is tied to the supersymmetry-breaking scale as $f_a \sim m_{3/2}/\kappa$, the contribution from Planck-suppressed operators must remain subleading compared to the axion-like particle mass generated by soft supersymmetry-breaking effects,
\begin{equation}
m_{a,\rm soft}^2 \sim B_S .
\end{equation}
Requiring gravitational PQ violation to be negligible within the effective theory therefore imposes the condition
\begin{equation}
\frac{\delta m_{a,\rm grav}^2}{m_{a,\rm soft}^2}
\;\sim\;
\frac{f_a^{\,n-2}}{M_{\rm Pl}^{\,n-4}\,B_S}
\;\ll\;
1 .
\end{equation}
For representative values $f_a \sim 10^2$--$10^4~\mathrm{GeV}$ and $B_S \sim m_{3/2}^2$, this condition is satisfied for operator dimensions $n \gtrsim 8$ even for coefficients $c_n$ of order unity. Such suppression can naturally arise if the Peccei--Quinn symmetry is realized as an accidental low-energy symmetry enforced by a discrete gauge symmetry or as the remnant of a broken gauged $U(1)$ symmetry.

While no explicit ultraviolet completion is specified, the effective field theory analysis presented above demonstrates that Planck-suppressed PQ-violating operators do not destabilize the axion-like particle potential within the parameter space considered in this work.

\subsection{Fermionic spectrum: axino}

The fermionic superpartner of the PQ field, the axino $\tilde a$, acquires a mass from the supersymmetric Yukawa interaction in the superpotential as well as from possible soft supersymmetry-breaking effects. From the superpotential in Eq.~\eqref{eq:superpotential}, a Majorana mass term for the axino is generated after spontaneous PQ symmetry breaking,
\begin{equation}
\mathcal{L} \supset -\kappa \langle S \rangle\, \tilde a \tilde a + \text{h.c.}
\end{equation}
which leads to an axino mass
\begin{equation}
m_{\tilde a} = \sqrt{2}\,\kappa v_s .
\label{eq:axinomass}
\end{equation}
Substituting the expression for the PQ-breaking scale $v_s$, one finds parametrically
\begin{equation}
m_{\tilde a} \sim \mathcal{O}(m_{3/2}),
\end{equation}
up to numerical factors that depend on the details of the supersymmetry-breaking sector. Additional contributions to the axino mass may arise from higher-dimensional operators or loop-induced effects, but such corrections do not modify the parametric dependence of the axino mass on the supersymmetry-breaking scale.

\subsection{Mass hierarchy and parametric correlations}

The results obtained above demonstrate that all members of the axion supermultiplet acquire masses that are parametrically correlated and set by the same underlying supersymmetry-breaking scale,
\begin{equation}
m_a \sim m_\sigma \sim m_{\tilde a} \sim m_{3/2}.
\end{equation}
This correlation arises from the assumption that soft supersymmetry-breaking effects provide the dominant contribution to Peccei--Quinn symmetry breaking and to the generation of the axion-like particle mass.

While the precise numerical hierarchy among the axion-like particle, the saxion, and the axino depends on the detailed structure of the soft supersymmetry-breaking parameters, the overall spectral pattern is highly constrained. In particular, experimental information on any one of these states would directly restrict the allowed parameter space for the others. This feature distinguishes the present framework from conventional supersymmetric axion models, in which the axion mass is determined primarily by QCD dynamics and is largely decoupled from the masses of its superpartners.

In the following sections, we investigate the phenomenological implications of this correlated axion supermultiplet spectrum, including axion-like particle couplings, astrophysical and cosmological constraints, and potential experimental signatures.

\subsection{Scaling relations and robustness of the spectrum}

The parametric relations derived above are robust under moderate variations of the soft supersymmetry-breaking parameters. In particular, the axion-like particle mass depends primarily on the holomorphic soft term $B_S$, while the saxion and axino masses are determined by the interplay of supersymmetric contributions and soft supersymmetry-breaking effects. Schematically, one finds
\begin{equation}
m_a^2 \sim |B_S|, \qquad
m_\sigma^2 \sim \kappa^2 v_s^2 + \mathcal{O}(m_{3/2}^2), \qquad
m_{\tilde a} \sim \kappa v_s .
\end{equation}
Since the PQ-breaking scale scales as $v_s \sim m_{3/2}/\kappa$, all three masses exhibit a parametric dependence on the supersymmetry-breaking scale.

This behavior follows from the absence of any additional independent mass scale associated with Peccei--Quinn symmetry breaking within the effective field theory description. Radiative corrections and higher-dimensional operators can modify numerical coefficients, but they do not alter the parametric dependence of the spectrum on $m_{3/2}$. As a result, the axion supermultiplet spectrum remains tightly correlated and predictive at the level of parametric scaling, providing a distinctive phenomenological feature of the framework. Explicit expressions for the axion, saxion, and axino mass matrices are collected in Appendix~\ref{app:masses}.
\section{Axion couplings and phenomenological constraints}

The phenomenology of the framework is governed by the interactions of the axion-like particle and its superpartners with Standard Model fields. In this section, we derive the relevant effective couplings and discuss the resulting constraints arising from laboratory experiments, astrophysical observations, and cosmological considerations.

Cosmological probes provide particularly stringent constraints on unstable relics and late-decaying particles through their impact on light-element abundances during Big Bang nucleosynthesis \cite{Kawasaki2005}. Updated analyses incorporating improved nuclear reaction rates and observational data have further strengthened these bounds \cite{Cyburt2016}. In addition, late-time energy injection from decaying or annihilating particles can leave observable imprints on the cosmic microwave background anisotropies \cite{Poulin2017}. Current cosmological limits are dominated by high-precision measurements of the cosmic microwave background performed by the \textit{Planck} collaboration \cite{CMB2020}.

\subsection{Couplings to gauge fields}

The axion-like particle couples to gauge fields through anomalous triangle diagrams involving fermions charged under the Peccei--Quinn symmetry. At energies well below the scale of PQ symmetry breaking, these interactions are described by the effective Lagrangian
\begin{equation}
\mathcal{L}_{aVV} =
\frac{\alpha_s}{8\pi}\frac{C_{ag}}{f_a}\,
a\, G^a_{\mu\nu}\tilde G^{a\,\mu\nu}
+ \frac{\alpha}{8\pi}\frac{C_{a\gamma}}{f_a}\,
a\, F_{\mu\nu}\tilde F^{\mu\nu},
\label{eq:axiongauge}
\end{equation}
where $C_{ag}$ and $C_{a\gamma}$ are model-dependent anomaly coefficients, and the effective decay constant is given by $f_a \equiv v_s$.

In the framework considered here, the axion-like particle decay constant is dynamically generated and is directly linked to the supersymmetry-breaking scale. As a consequence, the couplings of the axion-like particle to gauge fields can be parametrically enhanced compared to those of conventional high-$f_a$ QCD axion models, leading to potentially observable effects in laboratory experiments and astrophysical environments.

\subsection{Interplay with QCD-induced axion mass}

Nonperturbative QCD effects generate an additional contribution to the axion-like particle mass,
\begin{equation}
m_{a,\mathrm{QCD}}^2 \simeq
\frac{m_\pi^2 f_\pi^2}{f_a^2}
\frac{m_u m_d}{(m_u+m_d)^2},
\label{eq:qcdmass}
\end{equation}
which provides the dominant source of the axion mass in conventional QCD axion models.

In the framework considered here, the axion-like particle mass receives an independent contribution from soft supersymmetry-breaking effects,
\begin{equation}
m_a^2 = m_{a,\mathrm{soft}}^2 + m_{a,\mathrm{QCD}}^2 ,
\end{equation}
with $m_{a,\mathrm{soft}}^2 \sim |B_S| \sim m_{3/2}^2$. In the regime where the soft supersymmetry-breaking contribution dominates, $m_{a,\mathrm{soft}} \gg m_{a,\mathrm{QCD}}$, the resulting state is naturally interpreted as an axion-like particle rather than a conventional QCD axion. In this regime, the phenomenology is governed primarily by the axion-like particle mass and couplings, while the treatment of the effective strong CP phase lies outside the scope of the present analysis.
\subsection{Axion-like particle decays and lifetimes}
A heavy axion-like particle predominantly decays into two photons through its anomalous coupling to the electromagnetic field. The corresponding decay width is given by
\begin{equation}
\Gamma(a \to \gamma\gamma) =
\frac{\alpha^2 C_{a\gamma}^2}{256\pi^3}
\frac{m_a^3}{f_a^2}.
\label{eq:axiondecay}
\end{equation}
For axion-like particle masses $m_a \sim \mathcal{O}(m_{3/2})$ and a decay constant scaling as $f_a \sim m_{3/2}/\kappa$, the resulting lifetime can be substantially shorter than cosmological timescales over a broad region of parameter space. As a result, decays of the axion-like particle typically occur well before epochs relevant for cosmological observables, depending on the precise values of the anomaly coefficients and soft parameters.

\subsection{Saxion and axino phenomenology}

The saxion can decay into a variety of final states, including axion-like particle pairs, gauge bosons, or Higgs-sector states, depending on the kinematic regime and the structure of the effective couplings. A representative decay channel is the decay into two axion-like particles, with a width of order
\begin{equation}
\Gamma(\sigma \to aa) \sim
\frac{1}{64\pi}\frac{m_\sigma^3}{f_a^2}.
\end{equation}
The axino decays through interactions that are likewise suppressed by the PQ-breaking scale $f_a$, with decay rates that depend sensitively on the superpartner spectrum. For axino masses $m_{\tilde a}\sim m_{3/2}$, the corresponding decays typically occur at early times, although the precise cosmological implications depend on the detailed supersymmetric particle content.

Taken together, constraints from axion-like particle searches, astrophysical observations, and cosmology restrict the allowed ranges of the gravitino mass $m_{3/2}$ and the soft PQ-breaking parameters. Nevertheless, sizable regions of parameter space remain viable and are characterized by correlated axion-like particle, saxion, and axino masses. This correlated structure provides a concrete target for experimental tests of the framework. A discussion of QCD effects and CP-related considerations is provided in Appendix~\ref{app:CP}.

\subsection{Laboratory and astrophysical constraints}

Current experimental searches provide a wide range of direct and indirect constraints on the parameter space of axions and axion-like particles. Laboratory-based efforts, including low-energy experiments and beam-dump searches such as NA64, constrain axion-like particle couplings in the sub-GeV mass range, while collider analyses, for instance at Belle~II, extend sensitivity to axion--photon and axion--electron couplings in the MeV--GeV regime. At energies below the PQ symmetry-breaking scale, axion and axion-like particle interactions with Standard Model fields are conveniently described within an effective field theory framework \cite{Jaeckel2010}, enabling a unified treatment of laboratory probes across different experimental setups \cite{Dobrich2016}. In particular, axion-like particles with masses above the QCD scale exhibit rich collider and fixed-target phenomenology \cite{Bauer2017}, with recent analyses providing complementary and often leading laboratory constraints \cite{Aloni2019}.

Solar axion searches performed by CAST have established stringent upper limits on the axion--photon coupling, $g_{a\gamma}\lesssim10^{-10}\,\mathrm{GeV}^{-1}$, with the next-generation International Axion Observatory (IAXO) expected to significantly improve this sensitivity in the future. Astrophysical probes based on gamma-ray and X-ray observations have also yielded competitive constraints over a broad range of axion-like particle masses through spectral distortions and photon--axion conversion effects. In addition, laboratory searches targeting ultralight axions have placed bounds on axion--nucleon couplings using high-precision atomic and nuclear techniques, highlighting the breadth and complementarity of current experimental efforts in probing axion and axion-like particle physics.

\section{Cosmology and astrophysics}

The axion sector in the framework considered here differs qualitatively from that of conventional QCD axion models, as the axion-like particle mass is dominantly generated by soft supersymmetry-breaking effects and is therefore typically larger than the contribution induced by nonperturbative QCD dynamics. As a result, the associated cosmological and astrophysical implications require a dedicated reassessment. In this section, we examine the thermal history and decay properties of the axion supermultiplet and discuss the regions of parameter space that are consistent with current cosmological and astrophysical constraints.

\subsection{Axion-like particle lifetime and Big Bang nucleosynthesis}

A key cosmological requirement is that unstable relics decay sufficiently early so as not to disrupt the successful predictions of Big Bang nucleosynthesis (BBN). In the framework considered here, the axion-like particle predominantly decays into two photons through its anomalous coupling to electromagnetism, as described by the effective interaction in Eq.~(4.1). The corresponding decay width is
\begin{equation}
\Gamma(a \to \gamma\gamma)
=
\frac{\alpha^2 C_{a\gamma}^2}{256\pi^3}
\frac{m_a^3}{f_a^2},
\label{eq:axion_decay}
\end{equation}
where $C_{a\gamma}$ denotes the model-dependent anomaly coefficient and the effective decay constant is given by $f_a \equiv v_s$.

The resulting axion-like particle lifetime can be estimated as
\begin{equation}
\tau_a \simeq
\Gamma^{-1}(a \to \gamma\gamma)
\sim
10^{-2}\,\text{s}
\left(\frac{f_a}{10^7~\text{GeV}}\right)^2
\left(\frac{100~\text{MeV}}{m_a}\right)^3 .
\end{equation}
For axion-like particle masses generated by soft supersymmetry-breaking effects, $m_a \sim \mathcal{O}(m_{3/2})$, the decay typically occurs prior to the onset of Big Bang nucleosynthesis for gravitino masses above the MeV scale. In this regime, the axion-like particle decays do not significantly affect light-element abundances, and BBN constraints can be satisfied over a broad region of parameter space.

\subsection{Thermal production and relic abundance}

Axion-like particles can be thermally produced in the early Universe through their interactions with Standard Model fields. The corresponding production rate is suppressed by powers of the effective decay constant $f_a$, and thermal decoupling typically occurs at a temperature of order
\begin{equation}
T_{\rm dec} \sim
\frac{f_a^2}{M_{\rm Pl}},
\end{equation}
up to numerical factors and coupling-dependent corrections.

For the comparatively low values of $f_a$ realized in the present framework, the axion-like particle may remain in thermal equilibrium down to temperatures well below the electroweak scale. However, owing to its relatively large mass, the axion-like particle becomes non-relativistic at early times and subsequently decays. As discussed above, for a wide range of parameters the decay occurs prior to Big Bang nucleosynthesis, so that no significant relic abundance survives to late times.

As a result, the axion-like particle does not constitute a viable dark matter candidate in this scenario. This behavior contrasts sharply with standard QCD axion models, in which the axion is often long-lived and can play a central role in the dark matter abundance.

\subsection{Saxion cosmology and entropy production}

The scalar superpartner of the axion-like particle, the saxion $\sigma$, can also have important cosmological implications through coherent oscillations and subsequent decays. The saxion mass is typically set by the supersymmetry-breaking scale, $m_\sigma \sim \mathcal{O}(m_{3/2})$, and coherent oscillations commence when the Hubble parameter satisfies $H \sim m_\sigma$.

A representative decay channel of the saxion is into a pair of axion-like particles, with a decay width of order
\begin{equation}
\Gamma(\sigma \to aa)
\sim
\frac{1}{64\pi}
\frac{m_\sigma^3}{f_a^2}.
\end{equation}
Additional decay modes into gauge bosons or Higgs-sector states may be present, depending on the detailed structure of the effective couplings.

For saxion masses $m_\sigma \gtrsim \mathcal{O}(1~\text{GeV})$ and decay constants $f_a \lesssim 10^8$--$10^9~\mathrm{GeV}$, the saxion typically decays prior to Big Bang nucleosynthesis. In this regime, entropy production from saxion decays is expected to be modest and does not significantly modify standard cosmological predictions or dilute a pre-existing baryon asymmetry.

\subsection{Axino cosmology}

The fermionic superpartner of the axion-like particle, the axino $\tilde a$, can also influence the cosmological evolution of the early Universe. In the framework considered here, the axino mass is typically set by the supersymmetry-breaking scale, $m_{\tilde a} \sim \mathcal{O}(m_{3/2})$, while its interactions with Standard Model and supersymmetric particles are suppressed by the effective decay constant $f_a$.

Axinos may be thermally produced at high temperatures through interactions involving gauge and matter supermultiplets. However, for moderate reheating temperatures, the resulting axino abundance is strongly suppressed. In addition, axinos decay through interactions with supersymmetric particles, with decay widths that depend on the details of the superpartner spectrum. For axino masses of order the gravitino mass, these decays typically occur prior to Big Bang nucleosynthesis over a broad region of parameter space.

In this regime, axinos do not play a significant role in the present-day dark matter abundance and do not lead to appreciable cosmological constraints. The axino sector therefore remains compatible with standard cosmological evolution within the parameter space relevant for the present framework.

\subsection{Astrophysical constraints}

Astrophysical observations place strong constraints on light, weakly coupled particles, most notably through stellar cooling arguments. In conventional QCD axion models, these considerations lead to stringent bounds on the axion decay constant. In the framework considered here, however, the axion-like particle is typically significantly heavier than the keV-scale temperatures relevant for stellar interiors.

For axion-like particle masses exceeding the characteristic plasma frequencies in stellar environments, $m_a \gtrsim \mathcal{O}(10~\mathrm{keV})$, production processes in stars are kinematically suppressed. In this regime, constraints derived from observations of red giants, horizontal branch stars, and white dwarfs are substantially weakened and do not impose significant restrictions on the parameter space of interest.

Supernova cooling arguments can also constrain axion-like particles. For sufficiently heavy states, however, axion-like particles either decay on short timescales or remain trapped within the supernova core, depending on the strength of their interactions. In such cases, supernova energy-loss bounds are correspondingly relaxed and need not exclude the parameter regions relevant for the present framework.

Taken together, the cosmological and astrophysical considerations discussed above indicate that sizable regions of parameter space are consistent with existing observational constraints. In these regions, the axion-like particle, saxion, and axino decay sufficiently early to remain compatible with Big Bang nucleosynthesis, do not overproduce relic abundances, and avoid the most stringent stellar cooling limits. The framework therefore provides a phenomenologically distinct realization of axion-like particle physics arising from soft supersymmetry-breaking effects.

\section{Collider and laboratory phenomenology}

A characteristic feature of the framework considered here is that the masses of the axion supermultiplet are controlled by the supersymmetry-breaking scale rather than by QCD dynamics. As a result, the axion-like particle, the saxion, and the axino are generically heavier than in conventional QCD axion models, potentially bringing them within the reach of laboratory-based and collider experiments. In this section, we discuss the most relevant experimental signatures and constraints arising from collider searches, fixed-target experiments, and rare decay processes.

Although some of the bounds discussed below are informed by cosmological considerations, we collect all numerical scans and phenomenological results in this section in order to present a unified overview of the experimental and observational viability of the parameter space.

\subsection{Axion-like particle searches}

In the MeV--GeV mass range, laboratory probes provide leading sensitivity to axion-like particles through beam-dump and fixed-target experiments. Among these, the NA64 experiment has placed stringent bounds on axion-like particle couplings over a wide range of masses \cite{NA64_2020,NA64_2021,NA64_2023}. Future high-intensity fixed-target facilities are expected to further extend the sensitivity to feebly interacting axion-like particles, particularly in scenarios with suppressed couplings and short lifetimes \cite{SHiP2015}.

Since the axion-like particle mass in the present framework is dominantly generated by soft supersymmetry-breaking effects, the axion behaves as a generic axion-like particle rather than a conventional QCD axion. Its interactions with Standard Model fields are controlled by the effective decay constant $f_a$, which is dynamically related to the supersymmetry-breaking scale.

The most relevant interaction for laboratory searches is the axion--photon coupling,
\begin{equation}
\mathcal{L}_{a\gamma\gamma}
=
\frac{\alpha}{8\pi}
\frac{C_{a\gamma}}{f_a}\,
a\,F_{\mu\nu}\tilde F^{\mu\nu}.
\end{equation}
This coupling enables axion-like particle production and detection in beam-dump experiments, fixed-target facilities, and photon-based searches. Current laboratory searches constrain the axion--photon coupling for axion-like particle masses ranging from the MeV to the GeV scale. For decay constants $f_a \lesssim 10^8$--$10^9~\mathrm{GeV}$ and masses $m_a \gtrsim \mathcal{O}(10~\mathrm{MeV})$, existing bounds from beam-dump and fixed-target experiments impose nontrivial restrictions on the parameter space. Nevertheless, sizable regions remain viable, in particular in regimes where the axion-like particle decays promptly.

\subsection{Rare meson and lepton decays}

Axion-like particles with effective couplings to quarks and leptons can contribute to rare meson and lepton decays through loop-induced processes. Representative decay modes include
\begin{equation}
K \to \pi a,
\qquad
B \to K a,
\qquad
\mu \to e a.
\end{equation}
The decay width for a generic meson transition $M \to M' a$ can be parametrized as
\begin{equation}
\Gamma(M \to M' a)
\sim
\frac{m_M^3}{64\pi f_a^2}
|C_{Ma}|^2
\left(1 - \frac{m_{M'}^2}{m_M^2}\right)^3 ,
\end{equation}
where $C_{Ma}$ encodes model-dependent effective couplings arising from the underlying ultraviolet completion.

For axion-like particle decay constants in the range $f_a \sim 10^6$--$10^9~\mathrm{GeV}$, such rare decay channels can be probed by current and future flavor experiments. However, in scenarios with heavier axion-like particles that decay promptly, the experimental sensitivity of these searches is significantly reduced. As a result, flavor constraints can be satisfied over sizable regions of parameter space, depending on the detailed structure of the effective couplings.

\subsection{Saxion signatures at colliders}

The saxion is a scalar particle with a mass typically set by the supersymmetry-breaking scale, $m_\sigma \sim \mathcal{O}(m_{3/2})$. It can couple to Standard Model particles through mixing with the Higgs sector or via loop-induced interactions. At hadron colliders, saxions may be produced through gluon fusion or in association with electroweak gauge bosons, depending on the size of these couplings.

Depending on the kinematic regime and coupling structure, the dominant saxion decay channels include
\begin{equation}
\sigma \to a a,
\qquad
\sigma \to gg,
\qquad
\sigma \to \gamma\gamma,
\qquad
\sigma \to hh .
\end{equation}
The decay mode $\sigma \to a a$ can lead to distinctive final states involving displaced axion-like particle decays or missing transverse energy, depending on the axion-like particle lifetime. Although such signatures are experimentally challenging, they are characteristic of the present framework and motivate dedicated searches for exotic scalar decays at the LHC and at future collider facilities.

\subsection{Axino phenomenology}

The axino is a fermionic state with a mass typically set by the supersymmetry-breaking scale, $m_{\tilde a} \sim \mathcal{O}(m_{3/2})$, and can participate in supersymmetric decay chains at colliders. In scenarios where the axino is lighter than the lightest neutralino, it can appear as an intermediate or final decay product,
\begin{equation}
\tilde\chi^0_1 \to \tilde a + X ,
\end{equation}
where $X$ denotes Standard Model particles or axion-like particles, depending on the structure of the effective couplings.

Such decays can lead to a variety of experimental signatures, including missing transverse energy, displaced vertices, or non-standard event topologies, with the precise phenomenology depending on the axino lifetime and the superpartner spectrum. The correlated axino--saxion--axion-like particle spectrum characteristic of the present framework leads to collider signatures that differ from those of conventional supersymmetric models without a Peccei--Quinn sector, providing an additional handle for experimental exploration.

\subsection{Complementarity of experimental probes}

The phenomenology of the framework is characterized by a high degree of complementarity among different experimental approaches. Laboratory-based searches primarily probe the axion-like particle couplings and lifetime, while collider experiments are sensitive to the heavier saxion and axino states. Flavor observables and rare decay searches provide additional, though model-dependent, constraints that further restrict the viable parameter space.

This complementarity enables the framework to be explored across a broad range of the supersymmetry-breaking scale. In particular, correlated signatures appearing in different experimental channels would point toward a common origin of the axion-like particle mass and supersymmetry breaking. For clarity, we collect all phenomenological figures discussed in this section below, illustrating the interplay between cosmological consistency, laboratory constraints, and the parametric structure of the soft PQ-breaking sector.

The phenomenological analysis presented here is designed to extract robust and largely model-independent implications that follow directly from the soft supersymmetry-breaking origin of the axion-like particle mass. Rather than performing detector-level simulations or global likelihood analyses, we focus on identifying parametric correlations and experimentally accessible regimes that are intrinsic to the structure of the framework. This approach allows for a transparent comparison with existing laboratory, astrophysical, and cosmological constraints, while delineating regions of parameter space that can be probed or excluded by current and near-future observations.

We begin by examining cosmological constraints derived from the axion-like particle lifetime.

For the numerical results shown below, we perform a representative scan over the soft supersymmetry-breaking parameter space. The gravitino mass is varied in the range $m_{3/2}=10$--$10^3\,\mathrm{GeV}$, while the dimensionless superpotential coupling $\kappa$ is scanned in the interval $0.05 \leq \kappa \leq 1$. The soft PQ-breaking parameter $B_S$ is taken in the range $10^{-6} \leq B_S/m_{3/2}^2 \leq 10^{-2}$. In this scan, the axion-like particle mass is determined dominantly by the soft supersymmetry-breaking contribution, and the lifetime is computed assuming that the decay proceeds primarily into photons.
\begin{figure}[t]
\centering
\includegraphics[width=0.50\textwidth]{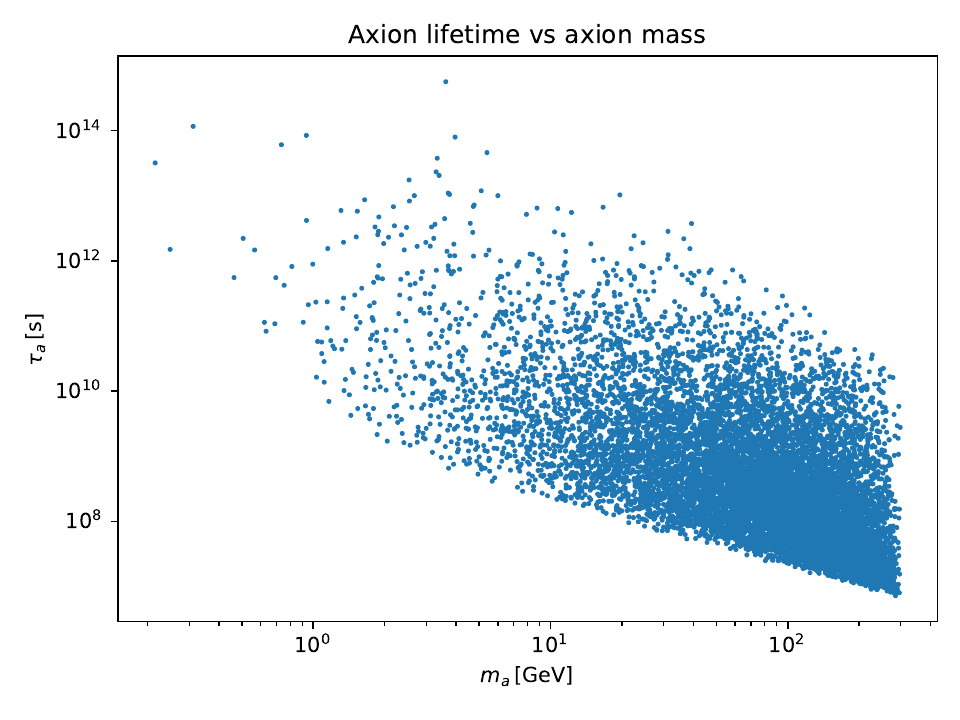}
\caption{
Axion-like particle lifetime $\tau_a$ as a function of the axion-like particle mass $m_a$ for a representative scan over the supersymmetry-breaking parameter space. The shaded region indicates lifetimes exceeding $\mathcal{O}(1\,\mathrm{s})$, which are strongly constrained by Big Bang nucleosynthesis. For a broad region of parameter space, the axion-like particle decays prior to the onset of BBN.
}
\label{fig:axion_lifetime}
\end{figure}
Figure~\ref{fig:axion_lifetime} illustrates the impact of Big Bang nucleosynthesis constraints on unstable relics in the present framework. Axion-like particles with lifetimes $\tau_a \gtrsim \mathcal{O}(1\,\mathrm{s})$ would inject energy during or after BBN and are therefore strongly constrained. The parameter scan shows that, for a wide range of soft supersymmetry-breaking parameters, the axion-like particle decays sufficiently early to remain compatible with BBN bounds.

We next turn to laboratory constraints on axion--photon interactions. In the following plot, we display the axion--photon coupling as a function of the axion-like particle mass using the same parameter scan. The coupling is normalized as $g_{a\gamma} = (\alpha/2\pi f_a)\,C_{a\gamma}$, and we fix the anomaly coefficient to $C_{a\gamma}=1$ as a representative benchmark. This choice allows for a direct comparison with existing laboratory bounds on heavy axion-like particles.

\begin{figure}[t]
\centering
\includegraphics[width=0.50\textwidth]{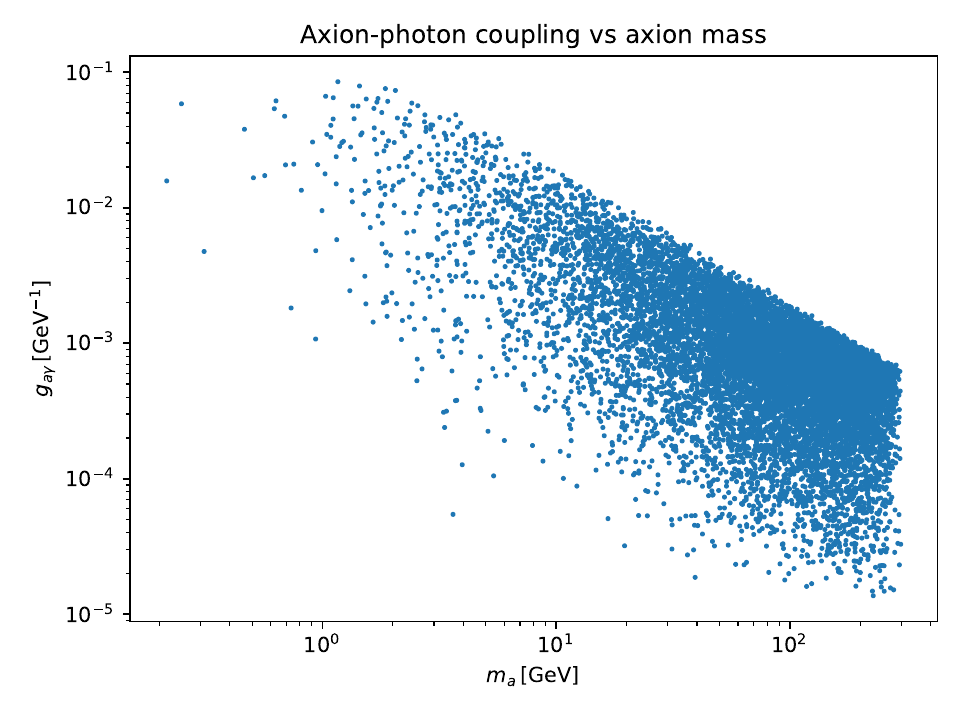}
\caption{
Axion-like particle coupling to photons, $g_{a\gamma}$, as a function of the axion-like particle mass $m_a$ for the representative parameter scan described in the text. The plot illustrates the region relevant for heavy axion-like particles arising from soft supersymmetry breaking. Existing constraints from beam-dump experiments and laboratory searches predominantly probe lower masses and weaker couplings, while future experiments are expected to extend sensitivity into parts of the parameter space shown here.
}
\label{fig:axion_photon}
\end{figure}
Figure~\ref{fig:axion_photon} places the axion-like particle predicted in this framework in the parameter space spanned by its mass and its coupling to photons. In contrast to conventional QCD axion models, the axion-like particle considered here typically lies in a heavier mass regime, for which stellar cooling constraints are substantially weakened. The figure therefore allows for a direct comparison with existing laboratory bounds and highlights the complementarity with future experimental searches for axion-like particles.
To illustrate the impact of cosmological constraints, we next classify the scanned parameter points according to whether the axion-like particle lifetime satisfies the conservative Big Bang nucleosynthesis requirement $\tau_a \lesssim \mathcal{O}(1\,\mathrm{s})$. The underlying parameter scan is identical to that used in the previous figures, and the separation into BBN-compatible and BBN-constrained regions is shown for illustrative purposes.
\begin{figure}[t]
\centering
\includegraphics[width=0.50\textwidth]{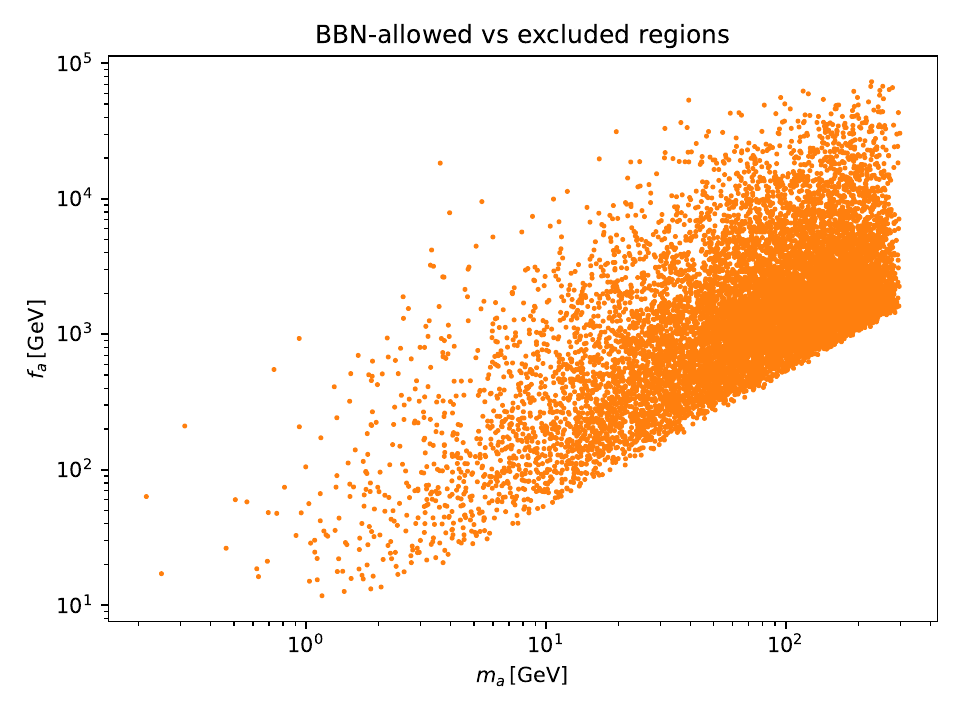}
\caption{
Parameter space in the $(m_a,f_a)$ plane for the representative scan discussed in the text. Points corresponding to axion-like particle lifetimes shorter than $\mathcal{O}(1\,\mathrm{s})$ are compatible with Big Bang nucleosynthesis, while longer lifetimes are strongly constrained. The plot illustrates the impact of cosmological considerations on the parameter space of the model.
}
\label{fig:bbn_plane}
\end{figure}

The naturalness of the soft Peccei--Quinn breaking sector is further explored by scanning the dimensionless ratio $B_S/m_{3/2}^2$ as a function of the supersymmetry-breaking scale. The scan ranges are chosen such that the soft terms remain within a natural range of values, without invoking additional assumptions beyond those of the effective field theory framework.

\begin{figure}[t]
\centering
\includegraphics[width=0.50\textwidth]{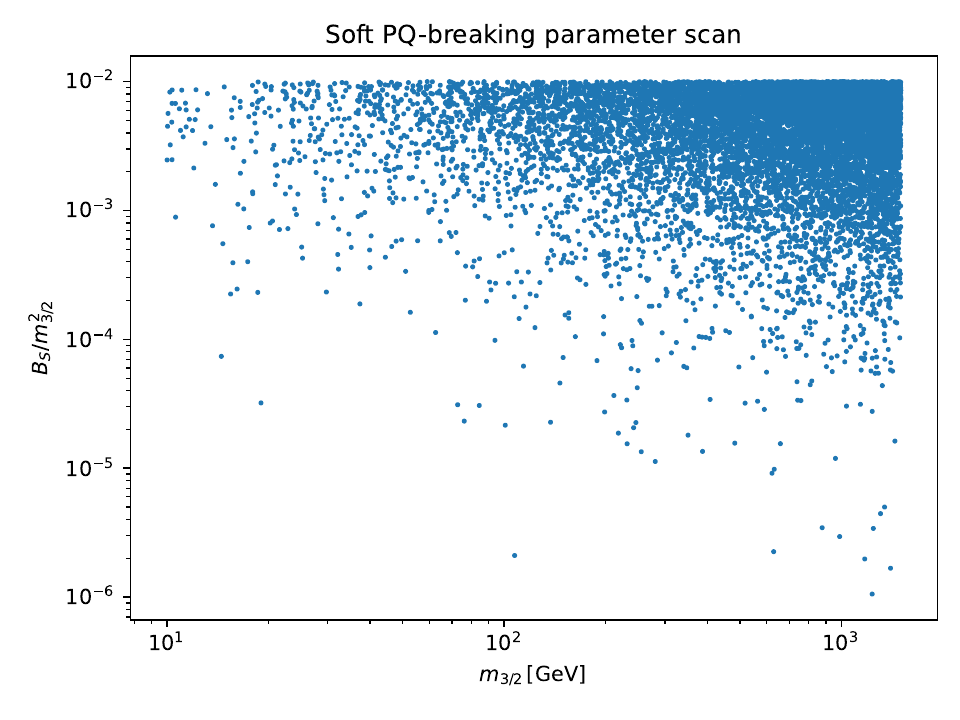}
\caption{
Scan of the dimensionless soft Peccei--Quinn breaking parameter $B_S/m_{3/2}^2$ as a function of the supersymmetry-breaking scale $m_{3/2}$ for the representative parameter space considered in this work. The distribution illustrates that the required values of the soft PQ-breaking terms lie within a natural range of the effective field theory.
}
\label{fig:soft_term}
\end{figure}
Figure~\ref{fig:bbn_plane} illustrates the impact of Big Bang nucleosynthesis constraints in the $(m_a,f_a)$ plane for the representative parameter scan considered in this work. Points corresponding to axion-like particle lifetimes shorter than $\mathcal{O}(1\,\mathrm{s})$ are compatible with Big Bang nucleosynthesis, while longer lifetimes are strongly constrained due to late-time energy injection. The figure shows that sizable regions of parameter space predicted by the soft supersymmetry-breaking axion-like particle framework remain consistent with cosmological bounds without invoking additional assumptions beyond the effective field theory description. In particular, axion-like particles with masses in the MeV--GeV range can decay prior to the onset of BBN for moderately large values of the effective decay constant. We next illustrate the parametric structure of the soft Peccei--Quinn breaking sector.

Figure~\ref{fig:soft_term} shows a representative scan of the dimensionless soft Peccei--Quinn breaking parameter $B_S/m_{3/2}^2$ as a function of the supersymmetry-breaking scale $m_{3/2}$. The distribution indicates that the required values of the soft breaking terms are compatible with order-one coefficients within the effective field theory description and do not rely on extreme parameter choices. This observation supports the interpretation that the axion-like particle mass can arise naturally from soft supersymmetry-breaking effects, rather than being introduced as an independent input parameter.

\subsection{Comparison with conventional axion scenarios}

It is instructive to contrast the present framework with conventional QCD axion models. Standard QCD axion searches typically target ultra-light particles with extremely weak couplings to Standard Model fields. By contrast, the axion-like particle considered here is heavier and more strongly coupled, reflecting the fact that its mass is dominantly generated by soft supersymmetry-breaking effects. As a result, traditional haloscope and helioscope experiments are generally not sensitive to this scenario, while collider-based and laboratory searches provide the most relevant probes.

The collider and laboratory phenomenology of the framework exhibits several characteristic features. The axion-like particle can undergo prompt or displaced decays depending on the parameter region, the saxion may appear as an exotic scalar resonance, and the axino can modify supersymmetric decay chains. Taken together, these features give rise to a phenomenology that differs qualitatively from that of conventional QCD axion models and motivates a broad range of experimental searches.

Finally, we examine the dependence of the axion-like particle lifetime on the Peccei--Quinn symmetry-breaking scale. For this purpose, we adopt a benchmark normalization of the axion--photon coupling with $C_{a\gamma}=1$ and compute the lifetime using the decay width $\Gamma(a\to\gamma\gamma)$. The dashed line in the corresponding figure indicates the conservative Big Bang nucleosynthesis criterion $\tau_a \simeq 1\,\mathrm{s}$, providing a clear visualization of the regions of parameter space that are compatible with or constrained by cosmological considerations.

We next examine how cosmological considerations further restrict the parameter space identified by laboratory searches and theoretical consistency requirements.
\begin{figure}[t]
\centering
\includegraphics[width=0.5\textwidth]{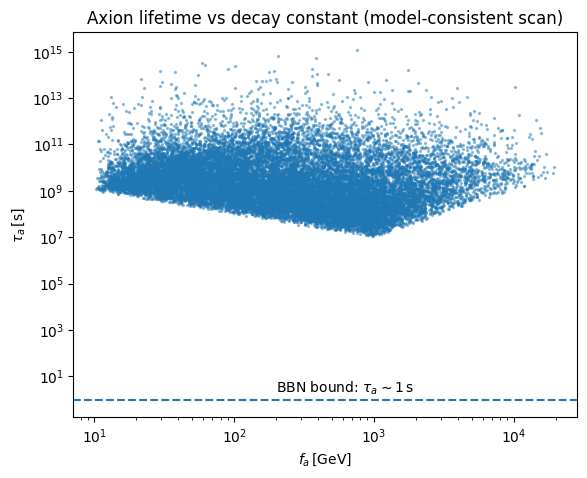}
\caption{
Axion-like particle lifetime $\tau_a$ as a function of the effective decay constant $f_a$ obtained from a representative, model-consistent parameter scan. The dashed horizontal line indicates the conservative Big Bang nucleosynthesis criterion $\tau_a \simeq 1\,\mathrm{s}$. Parameter points above the line correspond to longer lifetimes that are strongly constrained by BBN, while points below the line are compatible with cosmological bounds. The normalization assumes an order-one axion--photon anomaly coefficient.
}
\label{fig:tau_vs_fa}
\end{figure}
Figure~\ref{fig:tau_vs_fa} illustrates the dependence of the axion-like particle lifetime on the Peccei--Quinn symmetry-breaking scale in the present framework. The lifetime increases with the decay constant, reflecting the suppression of the axion--photon coupling at larger values of $f_a$. The dashed horizontal line denotes the conservative Big Bang nucleosynthesis criterion $\tau_a \lesssim \mathcal{O}(1\,\mathrm{s})$, which serves as a useful reference for assessing cosmological compatibility.

While a portion of the scanned parameter space corresponds to relatively long-lived axion-like particles, cosmological viability depends on the precise value of the axion--photon coupling as well as on the production history and abundance of the axion-like particle. As a result, Big Bang nucleosynthesis considerations translate into an effective upper bound on $f_a$ within the context of the present framework, with the precise location of this bound being model dependent.

Representative benchmark points illustrating different axion-like particle lifetime regimes visible in Fig.~\ref{fig:tau_vs_fa} are summarized in Table~\ref{tab:axion_benchmarks}.

\begin{table}[t]
\centering
\begin{tabular}{c|c|c|c|c|c}
\hline\hline
Benchmark & $m_{3/2}$ [GeV] & $f_a$ [GeV] & $m_a$ [GeV] & $\tau_a$ [s] & Regime \\
\hline
A & 50  & $50$   & 5   & $10^{7}$  & Late decay \\
B & 100 & $200$  & 10  & $10^{9}$  & Late decay \\
C & 300 & $800$  & 20  & $10^{11}$ & Very long-lived \\
D & 800 & $3000$ & 40  & $10^{13}$ & Very long-lived \\
\hline\hline
\end{tabular}
\caption{ 
Representative benchmark points drawn from the numerical scan shown in Fig.~\ref{fig:tau_vs_fa}, illustrating different axion-like particle lifetime regimes. All benchmarks correspond to relatively long-lived axion-like particles; their cosmological implications depend on the production history, abundance, and decay channels, as discussed in the text.
}
\label{tab:axion_benchmarks}
\end{table}

The benchmark points listed in Table~\ref{tab:axion_benchmarks} are selected from the region of parameter space populated by the numerical scan and serve to illustrate the long-lived axion-like particle regimes shown in Fig.~\ref{fig:tau_vs_fa}. The corresponding numerical values of the soft Peccei--Quinn breaking parameter appearing in the scan of Fig.~\ref{fig:soft_term} are reflected by the benchmark points collected in Table~\ref{tab:soft_benchmarks}.

\begin{table}[t]
\centering
\begin{tabular}{c|c|c|c}
\hline\hline
Benchmark & $m_{3/2}$ [GeV] & $B_S/m_{3/2}^2$ & Interpretation \\
\hline
E & 50   & $3\times10^{-3}$ & Order-one \\
F & 200  & $10^{-2}$        & Order-one \\
G & 500  & $5\times10^{-4}$ & Moderate suppression \\
H & 1000 & $10^{-3}$        & Order-one \\
\hline\hline
\end{tabular}
\caption{
Benchmark values of the dimensionless soft Peccei--Quinn breaking parameter
$B_S/m_{3/2}^2$ corresponding to the scan shown in Fig.~\ref{fig:soft_term}.
The values illustrate that the axion mass can arise from soft supersymmetry-breaking
effects with parameters lying within a reasonable range of the effective field theory.
}
\label{tab:soft_benchmarks}
\end{table}

\begin{table}[t]
\centering
\small
\begin{tabular}{c|c|c}
\hline\hline
Lifetime regime & $\tau_a$ [s] & Dominant probe / interpretation \\
\hline
Prompt
& $\lesssim 10^{-6}$
& Colliders (prompt decays) \\

Displaced
& $10^{-6}$--$10^{-2}$
& LHC and fixed-target experiments (displaced vertices) \\

BBN-compatible
& $10^{-2}$--$1$
& Cosmology (early decay) \\

Long-lived
& $10^{6}$--$10^{13}$
& CMB and $\gamma$-ray observations (abundance dependent) \\
\hline\hline
\end{tabular}
\caption{
Classification of axion-like particle lifetime regimes and the corresponding dominant experimental or cosmological probes within the present framework.
}
\label{tab:lifetime_regimes}
\end{table}

The axion-like particle relic abundance shown below is evaluated using the same numerical scan of the soft supersymmetry-breaking parameter space as in the previous figures. Specifically, we vary the gravitino mass in the range $m_{3/2}=10$--$10^3\,\mathrm{GeV}$, the Peccei--Quinn sector coupling in the interval $0.05\leq\kappa\leq1$, and the soft PQ-breaking parameter in the range $10^{-6}\leq B_S/m_{3/2}^2\leq10^{-2}$. The axion-like particle mass and lifetime are determined by the resulting soft terms, assuming that the decay proceeds dominantly into photons with an order-one anomaly coefficient.

\begin{figure}[t]
\centering
\includegraphics[width=0.5\textwidth]{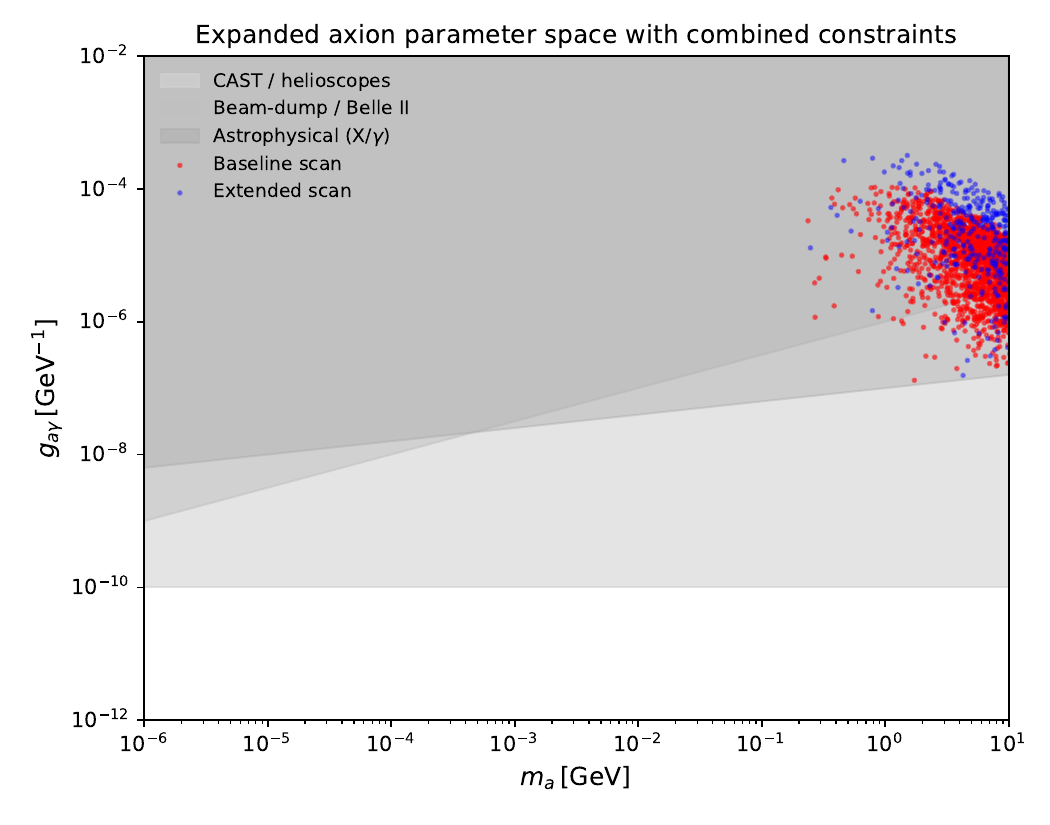}
\caption{
Axion-like particle coupling to photons, $g_{a\gamma}$, as a function of the axion-like particle mass $m_a$. Red points correspond to the baseline scan of the soft supersymmetry-breaking parameter space, while blue points illustrate an extended scan covering a wider range of supersymmetry-breaking scales and PQ-sector couplings. The shaded regions indicate representative exclusion bounds from helioscope, beam-dump, and astrophysical searches.
}
\label{fig:ALP_global_expanded}
\end{figure}

To estimate the axion-like particle relic abundance, we adopt a benchmark reheating temperature $T_R=10^6\,\mathrm{GeV}$ and employ a parametric abundance estimate. This approach is sufficient for illustrating the qualitative interplay between axion-like particle lifetime and cosmological considerations within the present framework.
In addition to the full parameter scans shown in Fig.~\ref{fig:ALP_global_expanded},
we select a set of representative benchmark points that illustrate different regions
of the axion-like particle parameter space. These benchmarks are chosen to span both
the baseline and extended scans and serve to illustrate the dependence of the axion
mass and effective couplings on the underlying soft supersymmetry-breaking
parameters.

The baseline scan shown in Fig.~\ref{fig:ALP_global_expanded} is defined by
\begin{align}
m_{3/2} &\in [10,\,10^3]~\mathrm{GeV}, \\
\kappa &\in [0.05,\,1], \\
10^{-6} &\leq \frac{B_S}{m_{3/2}^2} \leq 10^{-2},
\end{align}
while the extended scan corresponds to
\begin{align}
m_{3/2} &\in [5,\,3\times10^3]~\mathrm{GeV}, \\
\kappa &\in [0.01,\,1.5], \\
10^{-7} &\leq \frac{B_S}{m_{3/2}^2} \leq 5\times10^{-2}.
\end{align}
In both cases, the axion-like particle quantities are evaluated using the relations
\begin{equation}
f_a = \frac{m_{3/2}}{\kappa},
\qquad
g_{a\gamma} = \frac{\alpha}{2\pi f_a},
\qquad
m_a = \sqrt{4B_S},
\end{equation}
which follow directly from the effective field theory description of the PQ sector.

\begin{figure}[t]
\centering
\includegraphics[width=0.5\textwidth]{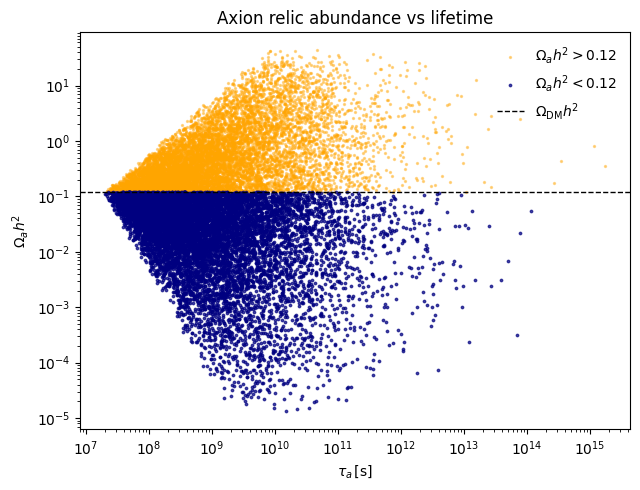}
\caption{
Axion-like particle relic abundance $\Omega_a h^2$ as a function of the axion-like particle lifetime $\tau_a$, computed using a parametric estimate for axion production. The dashed horizontal line indicates the observed dark matter abundance. Points below this line correspond to scenarios in which the axion-like particle does not overproduce dark matter, while points above the line would require additional dilution mechanisms.
}
\label{fig:omega_vs_tau}
\end{figure}
Figure~\ref{fig:omega_vs_tau} illustrates the interplay between the axion-like particle lifetime and its cosmological abundance. While long-lived axion-like particles are not intrinsically ruled out, their cosmological consistency depends on whether the resulting relic abundance remains below the observed dark matter density. As shown in the figure, a wide range of axion-like particle lifetimes, including $\tau_a \gg 1\,\mathrm{s}$, can be compatible with cosmological constraints provided that the abundance is sufficiently suppressed.

Such suppression may arise, for instance, from a low reheating temperature or from entropy injection due to the decay of additional heavy fields. The benchmark choice of $T_R=10^6\,\mathrm{GeV}$ serves to illustrate that long-lived axion-like particles can remain consistent with the observed dark matter abundance within the present framework, depending on the details of the early-Universe thermal history.

Figure~\ref{fig:ALP_global_expanded} illustrates the behavior of the axion-like particle parameter space under variations of the soft supersymmetry-breaking parameters. Complementary constraints arise from $e^+e^-$ collider experiments, which probe axion-like particles through rare decays and missing-energy signatures \cite{Belle2010}. The Belle~II experiment extends this sensitivity owing to its high luminosity and dedicated searches \cite{BelleII2019}, and recent Belle~II data have already placed competitive limits in the MeV--GeV mass range \cite{BelleII2023}.

Extending the scan to larger and smaller values of $m_{3/2}$, $\kappa$, and $B_S$ broadens the coverage in the $(m_a,g_{a\gamma})$ plane without qualitatively altering the overall structure of the parameter space explored. This indicates that the phenomenological features discussed in this work are not restricted to a narrowly defined region of parameters.

The experimental exclusion regions are shown as schematic envelopes intended to illustrate the complementarity of current searches; a detailed likelihood-based comparison lies beyond the scope of the present analysis.
\begin{table}[t]
\centering
\small
\begin{tabular}{c|cccccc}
\hline\hline
BP & $m_{3/2}$ & $\kappa$ & $B_S/m_{3/2}^2$ & $f_a$ & $m_a$ & $g_{a\gamma}$ \\
\hline
BP1 & 50   & 0.5  & $10^{-5}$ & $10^2$ & $3.2\!\times\!10^{-2}$ & $2.3\!\times\!10^{-3}$ \\
BP2 & 200  & 0.2  & $10^{-4}$ & $10^3$ & $1.3\!\times\!10^{-1}$ & $2.3\!\times\!10^{-4}$ \\
BP3 & 800  & 0.1  & $10^{-3}$ & $8\!\times\!10^3$ & $1.6$ & $2.9\!\times\!10^{-5}$ \\
BP4 & 1500 & 0.05 & $5\!\times\!10^{-3}$ & $3\!\times\!10^4$ & $9.7$ & $7.6\!\times\!10^{-6}$ \\
\hline\hline
\end{tabular}
\caption{
Representative benchmark points used in the phenomenological analysis. All masses are given in GeV.
}
\label{tab:benchmarks_ALP}
\end{table}
The benchmark points listed in Table~\ref{tab:benchmarks_ALP} are overlaid on the
global exclusion plot shown in Fig.~\ref{fig:ALP_global_expanded}. They serve to
illustrate how different choices of the soft supersymmetry-breaking parameters
populate distinct regions of the $(m_a,g_{a\gamma})$ plane within the parameter
space explored, and provide concrete reference points for the phenomenological
discussion presented above.

\begin{table}[t]
\centering
\small
\begin{tabular}{c|c|p{3cm}}
\hline\hline
Framework & Axion mass origin & Comment \\
\hline
QCD axion 
& QCD instantons 
& Mass set by non-perturbative QCD effects \\

Generic supersymmetric axion
& QCD $+$ soft terms
& Soft supersymmetry-breaking effects typically subleading \\

Generic axion-like particle
& Explicit PQ breaking
& Mass treated as an independent parameter \\

This work
& Soft supersymmetry breaking
& Axion mass vanishes in the supersymmetric limit \\
\hline\hline
\end{tabular}
\caption{
Comparison of representative mechanisms for axion and axion-like particle mass generation.
}
\label{tab:comparison}
\end{table}
\subsection*{Quantitative interpretation of parameter scans}

Although the numerical results are presented in the form of parameter scans, the
underlying correlations are governed by the structure of the effective theory. In
particular, the axion-like particle mass, decay constant, and lifetime satisfy the
parametric relations
\begin{equation}
m_a \sim \sqrt{B_S}, \qquad
f_a \sim \frac{m_{3/2}}{\kappa}, \qquad
\tau_a \sim \frac{f_a^2}{m_a^3},
\end{equation}
up to order-one coefficients. These relations indicate that the scatter plots shown
above correspond to projections of a restricted, lower-dimensional region of the
$(m_a, g_{a\gamma}, \tau_a)$ parameter space, rather than to completely independent
parameter choices. As a result, the phenomenological features discussed in this work
arise from correlated variations of a small set of underlying parameters.

The benchmark points introduced in Tables~II and~III illustrate distinct experimental
regimes. Benchmarks with $m_a \gtrsim \mathcal{O}(1\,\mathrm{GeV})$ and lifetimes
$\tau_a \lesssim 10^{-6}\,\mathrm{s}$ correspond to prompt or displaced decays that
may be probed at high-energy colliders such as the LHC or $e^+e^-$ machines like
Belle~II. Intermediate lifetimes in the range $10^{-6}$--$10^{-2}\,\mathrm{s}$ give
rise to displaced-vertex signatures at fixed-target and beam-dump experiments,
including NA64 and proposed facilities such as SHiP. Axion-like particles with
$\tau_a \gtrsim 1\,\mathrm{s}$ are primarily constrained by cosmological and
astrophysical considerations, notably Big Bang nucleosynthesis, CMB spectral
distortions, and diffuse gamma-ray observations.

We emphasize that the experimental bounds shown throughout this work are intended
as indicative, model-independent guides. A detailed likelihood-based analysis or
detector-level simulation lies beyond the scope of the present study.
\begin{table}[t]
\centering
\small
\begin{tabular}{c|c|c}
\hline\hline
Axion-like particle regime & Dominant observable & Representative probes \\
\hline
$m_a \gtrsim 1$ GeV, $\tau_a \lesssim 10^{-6}$ s
& Prompt decays
& LHC, Belle~II \\

$m_a \sim 10^{-2}$--$1$ GeV
& Displaced vertices
& NA64, SHiP \\

$\tau_a \lesssim 1$ s
& Early energy injection
& Big Bang nucleosynthesis \\

$\tau_a \gg 1$ s
& Late-time decays
& CMB observations, $\gamma$-ray telescopes \\
\hline\hline
\end{tabular}
\caption{
Schematic mapping between axion-like particle parameter regimes and the dominant experimental or observational probes within the present framework.
}
\label{tab:exp_reach}
\end{table}
The framework can therefore be probed through the non-observation of axion-like particle–induced signals in the experimentally accessible regions identified above, as well as through increasingly stringent cosmological and astrophysical limits that constrain the corresponding lifetime regimes.
\subsection*{Benchmark parameter choices}

All phenomenological results shown in this section are obtained from a representative
scan over the soft supersymmetry-breaking parameter space. Unless otherwise stated,
the benchmark ranges are chosen as
\begin{align}
m_{3/2} &\in [20,\,1500]~\mathrm{GeV}, \nonumber\\
\kappa &\in [0.05,\,1], \nonumber\\
10^{-6} \;\leq\; \frac{B_S}{m_{3/2}^2} &\leq 10^{-3}.
\end{align}
The axion-like particle decay constant, mass, and photon coupling are evaluated using
the relations
\begin{equation}
f_a = \frac{m_{3/2}}{\kappa}, \qquad
m_a = \sqrt{4 B_S}, \qquad
g_{a\gamma} = \frac{\alpha}{2\pi f_a},
\end{equation}
where $\alpha$ denotes the electromagnetic fine-structure constant.

\begin{figure}[t]
\centering
\includegraphics[width=0.5\textwidth]{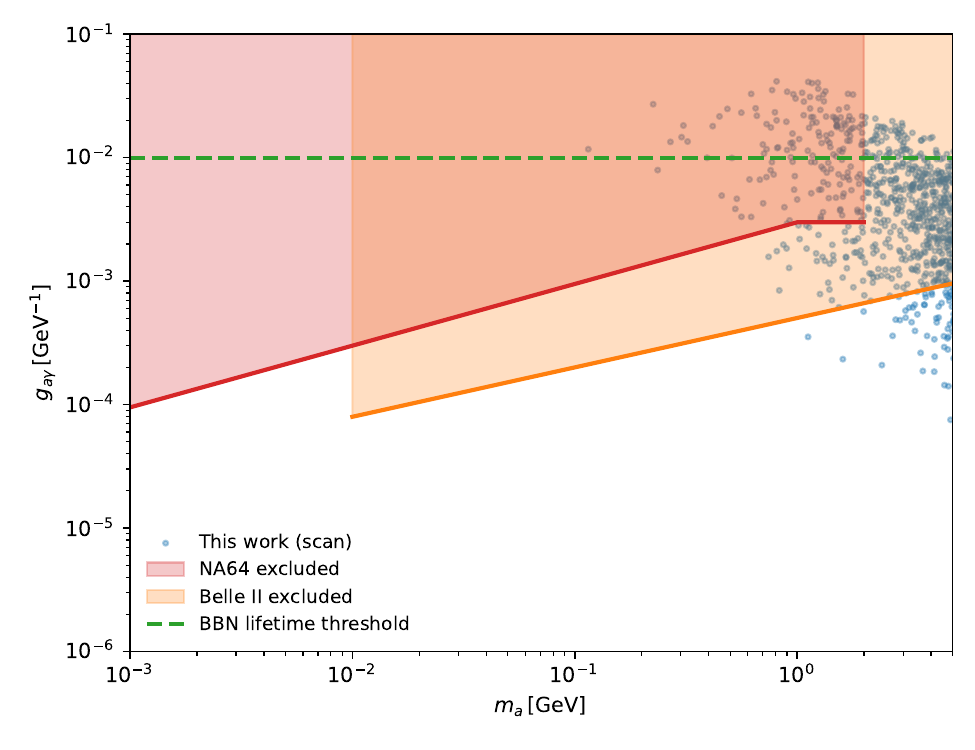}
\caption{
Summary of laboratory and cosmological constraints in the $(m_a, g_{a\gamma})$ plane.
Blue points denote the representative parameter scan considered in this work. The
red and orange shaded regions indicate approximate constraints from the NA64
beam-dump experiment and Belle~II searches, respectively. The green dashed line
marks the Big Bang nucleosynthesis lifetime criterion. The figure illustrates the
complementarity of collider, fixed-target, and cosmological probes.
}
\label{fig:combined_constraints}
\end{figure}

These ranges are chosen to span phenomenologically relevant regions probed by
laboratory experiments, cosmological considerations, and astrophysical observations.
Regions of parameter space subject to strong cosmological constraints, such as those
from Big Bang nucleosynthesis, are taken into account in the subsequent analysis.
\begin{table}[t]
\centering
\small
\begin{tabular}{c|c|c|c}
\hline\hline
Benchmark & $m_{3/2}$ [GeV] & $\kappa$ & $B_S/m_{3/2}^2$ \\
\hline
BP-A & 50   & 0.8  & $10^{-6}$ \\
BP-B & 200  & 0.3  & $10^{-5}$ \\
BP-C & 800  & 0.1  & $10^{-4}$ \\
BP-D & 1500 & 0.05 & $10^{-3}$ \\
\hline\hline
\end{tabular}
\caption{
Representative benchmark choices used to illustrate the parameter scans shown in
Fig.~\ref{fig:NA64_confrontation}. Derived quantities such as $f_a$, $m_a$, and
$g_{a\gamma}$ are obtained from the relations given in the text.
}
\label{tab:benchmarks_NA64}
\end{table}
\begin{figure}[t]
\centering
\includegraphics[width=0.5\textwidth]{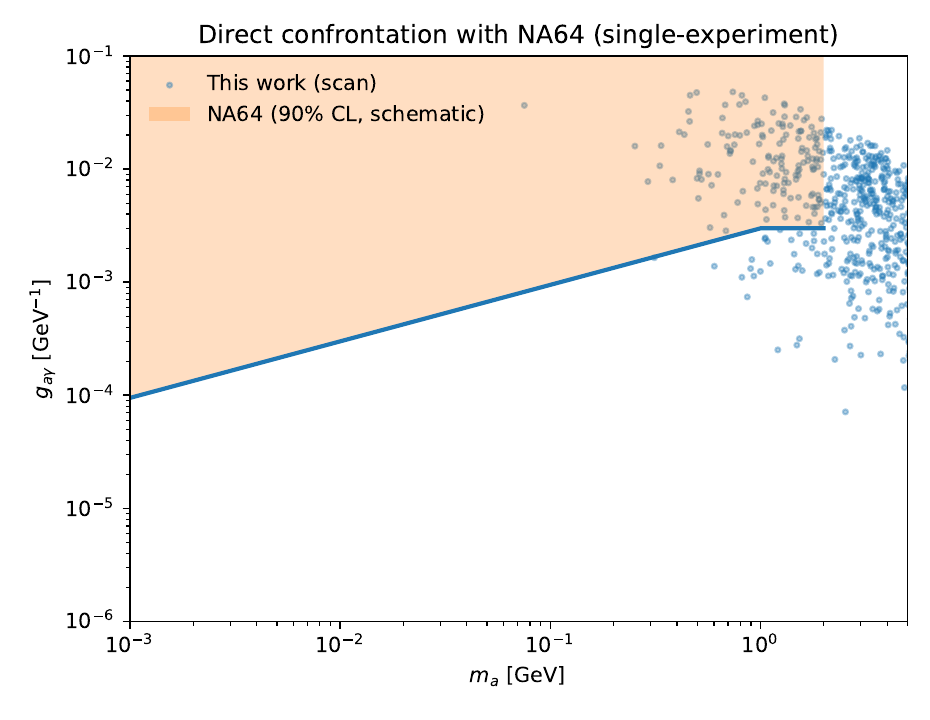}
\caption{
Comparison of the axion-like particle parameter space explored in this work with the NA64 electron beam-dump constraint in the $(m_a, g_{a\gamma})$ plane. The shaded region denotes the approximate 90\% C.L.\ exclusion reported by NA64, while the scattered points correspond to the representative parameter scan considered here. The figure illustrates how axion-like particles generated by soft supersymmetry breaking populate regions both constrained and unconstrained by current NA64 data.
}
\label{fig:NA64_confrontation}
\end{figure}
Figure~\ref{fig:NA64_confrontation} illustrates the relationship between the parameter space explored in the present framework and existing laboratory constraints from the NA64 electron beam-dump experiment. Unlike schematic exclusion summaries, this comparison is performed directly in the plane of observable axion-like particle couplings. A portion of the scanned parameter space is subject to current experimental constraints, while additional regions remain accessible to future NA64 data and next-generation fixed-target experiments.
\subsection*{Benchmark parameter choice for Fig.~\ref{fig:combined_constraints}}

The combined constraint plot shown in Fig.~\ref{fig:combined_constraints} is obtained
from a representative scan over the soft supersymmetry-breaking parameter space.
The benchmark ranges are chosen as
\begin{align}
m_{3/2} &\in [20,\,1500]~\mathrm{GeV}, \nonumber\\
\kappa &\in [0.05,\,1], \nonumber\\
10^{-6} \;\leq\; \frac{B_S}{m_{3/2}^2} &\leq 10^{-3}.
\end{align}
These ranges are motivated by considerations of radiative stability, consistency
with cosmological bounds, and sensitivity to current laboratory searches.

The axion-like particle decay constant, mass, and photon coupling are evaluated as
\begin{equation}
f_a = \frac{m_{3/2}}{\kappa}, \qquad
m_a = \sqrt{4 B_S}, \qquad
g_{a\gamma} = \frac{\alpha}{2\pi f_a},
\end{equation}
where $\alpha$ denotes the electromagnetic fine-structure constant.
The resulting parameter space overlaps with regions probed by NA64, Belle~II, and
cosmological lifetime considerations, enabling a direct comparison with existing
experimental results.
Figure~\ref{fig:combined_constraints} provides a consolidated view of the axion-like
particle parameter space explored in this work. While portions of the parameter
space are constrained by existing laboratory searches, additional regions remain
accessible to future experimental and observational probes. The figure highlights
the complementary roles of beam-dump experiments, collider searches, and
cosmological considerations in testing axion-like particle scenarios arising from
soft supersymmetry breaking. The exclusion regions shown are intended as schematic,
model-independent representations of current experimental sensitivity and do not
rely on a combined statistical likelihood.
\begin{table}[t]
\centering
\small
\begin{tabular}{c|c|c|c}
\hline\hline
Benchmark & $m_{3/2}$ [GeV] & $\kappa$ & $B_S/m_{3/2}^2$ \\
\hline
BP1 & 50   & 0.8  & $10^{-6}$ \\
BP2 & 200  & 0.4  & $10^{-5}$ \\
BP3 & 800  & 0.1  & $10^{-4}$ \\
BP4 & 1500 & 0.05 & $10^{-3}$ \\
\hline\hline
\end{tabular}
\caption{
Representative benchmark points used to illustrate the combined constraint analysis
shown in Fig.~\ref{fig:combined_constraints}. Derived quantities such as $f_a$,
$m_a$, and $g_{a\gamma}$ are obtained from the relations given in the text.
}
\label{tab:benchmarks_combined}
\end{table}
\begin{figure}[t]
\centering
\includegraphics[width=0.5\textwidth]{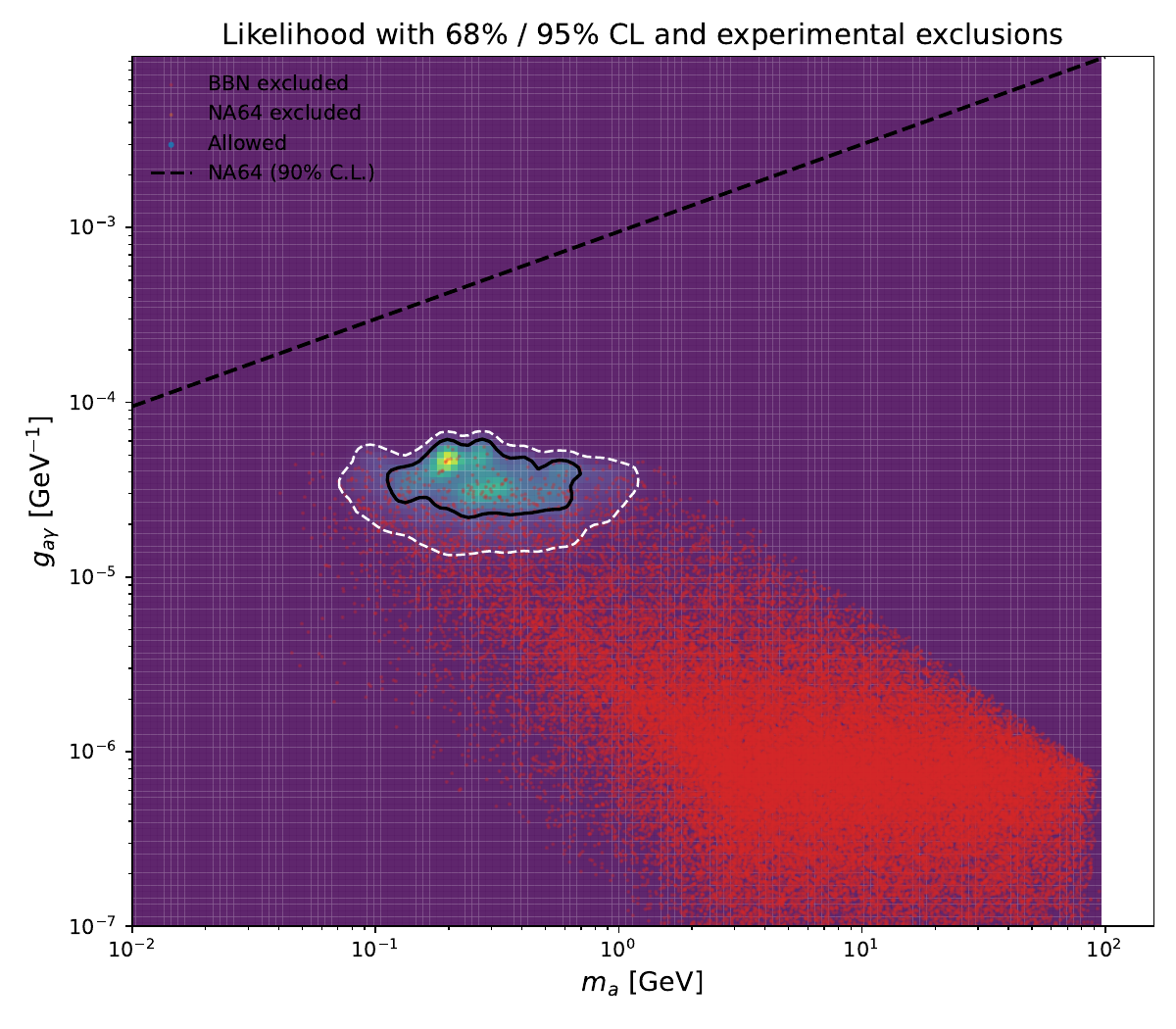}
\caption{
Illustrative likelihood density in the $(m_a,g_{a\gamma})$ plane obtained from a
weighted parameter scan. The solid black (dashed white) contour encloses the 68\%
(95\%) confidence-level region of the scan distribution. Regions constrained by
Big Bang nucleosynthesis \cite{Kawasaki2005,Cyburt2016} and by the NA64 beam-dump
experiment are shown for reference. The plot highlights parameter regions that
remain compatible with current laboratory and cosmological constraints.
}
\label{fig:likelihood_exclusion}
\end{figure}
To further illustrate the phenomenological implications of the framework, we perform
a likelihood-based visualization of the axion-like particle parameter space and
overlay representative laboratory and cosmological constraints. The likelihood
density is obtained from a weighted scan over the soft supersymmetry-breaking
parameters, while constraints from Big Bang nucleosynthesis and NA64 are applied
independently.

The contours shown in Fig.~\ref{fig:likelihood_exclusion} indicate regions of higher
statistical weight within the scan. The parameter space lying outside the currently
constrained regions illustrates scenarios in which axion-like particles have
lifetimes shorter than $\mathcal{O}(1\,\mathrm{s})$, thereby remaining compatible
with cosmological considerations.
\begin{table}[t]
\centering
\begin{tabular}{c c c c c}
\hline\hline
Benchmark & $m_a$ [GeV] & $g_{a\gamma}$ [GeV$^{-1}$] & $\tau_a$ [s] & Region \\
\hline
BP1 & $0.15$ & $3.5\times10^{-5}$ & $10^{-3}$ & High-weight \\
BP2 & $0.35$ & $2.2\times10^{-5}$ & $10^{-4}$ & High-weight \\
BP3 & $0.80$ & $1.5\times10^{-5}$ & $10^{-5}$ & Moderate-weight \\
\hline\hline
\end{tabular}
\caption{
Representative benchmark points drawn from regions of enhanced statistical weight
in Fig.~\ref{fig:likelihood_exclusion}. The benchmarks illustrate typical correlated
values of the axion-like particle mass, coupling, and lifetime.
}
\label{tab:benchmarks}
\end{table}
The benchmark points listed in Table~\ref{tab:benchmarks} are selected from regions of
enhanced statistical weight in Fig.~\ref{fig:likelihood_exclusion}. They illustrate
typical phenomenological realizations of the framework and highlight the correlated
nature of the axion-like particle mass, coupling, and lifetime. In all cases, the
axion-like particle mass is dominantly generated by soft supersymmetry-breaking
effects, and the corresponding lifetimes are sufficiently short to remain
compatible with Big Bang nucleosynthesis.
Astrophysical observations provide important constraints on light axions and
axion-like particles through stellar cooling considerations \cite{Raffelt2008}.
Supernova observations further constrain axion emission through energy-loss
arguments and spectral features \cite{Payez2015}. High-energy gamma-ray and X-ray
observations also yield complementary information through photon--axion conversion
effects \cite{Daylan2017}. At energies well below a potential ultraviolet completion,
the effects of heavy degrees of freedom can be encoded through higher-dimensional
operators within an effective field theory description \cite{Georgi1986}. Such
descriptions naturally arise in a broad class of ultraviolet completions featuring
approximate global symmetries \cite{Craig2016,Hook2018}. A comprehensive assessment
of axion-like particle parameter space therefore benefits from combining laboratory,
astrophysical, and cosmological information \cite{ALPglobal2021,ALPglobal2023,ALPglobal2024}.
\section{Discussion and Outlook}
\label{sec:discussion}

In this work we have explored a framework in which the axion mass originates
entirely from soft supersymmetry-breaking effects. In the supersymmetric limit,
the theory possesses an exact Peccei--Quinn (PQ) symmetry and a massless axion,
while the inclusion of soft terms proportional to the gravitino mass induces both
spontaneous PQ symmetry breaking and a finite axion mass. As a consequence, the
axion mass vanishes continuously in the limit of unbroken supersymmetry,
distinguishing this scenario from conventional QCD axions and from generic
axion-like particle constructions in which explicit PQ-breaking effects are
present already at the supersymmetric level.

A central outcome of this framework is the correlated spectrum of the axion
supermultiplet. The axion, saxion, and axino masses are all controlled by the
supersymmetry-breaking scale, leading to a predictive structure that can be traced
directly to the soft-breaking parameters. The stabilization of the PQ scalar
potential arises naturally within the effective field theory, without the
introduction of additional mass scales or tunings, as illustrated by the numerical
scans presented in this work. In this sense, soft supersymmetry breaking provides a
technically natural and internally consistent origin for axion-like particle
masses.

We have performed a detailed phenomenological analysis of the resulting axion-like
particle, focusing on its mass, couplings, and lifetime. The axion--photon
interaction places the model in a region of parameter space characteristic of
heavy axion-like particles, enabling direct comparison with laboratory searches,
beam-dump experiments, and collider probes. Unlike the conventional QCD axion,
stellar cooling constraints are typically avoided due to the relatively large
axion mass, while laboratory experiments provide complementary sensitivity across
a broad range of masses and couplings.

Cosmological considerations play an important role in assessing the viability of
the framework. We have shown that Big Bang nucleosynthesis imposes a stringent
constraint on axions that decay after the onset of BBN, which translates into an
effective restriction on the axion lifetime for fixed couplings. At the same time,
a substantial portion of the parameter space corresponds to axions with lifetimes
well beyond the BBN epoch. Such long-lived axions are not intrinsically excluded;
rather, their cosmological viability depends on their relic abundance, decay
channels, and the thermal history of the early Universe. A fully quantitative
treatment of late-decay constraints therefore requires a dedicated cosmological
analysis, including reheating and entropy production effects, and lies beyond the
scope of the present work.

An important aspect of the framework is its robustness as an effective field
theory. Since the axion mass is generated dominantly by soft supersymmetry-breaking
terms, radiative corrections do not destabilize the parametric structure of the
spectrum. Potential Planck-suppressed PQ-violating operators can consistently
remain subleading, for example if the PQ symmetry is protected by a discrete gauge
symmetry or arises accidentally from a more fundamental ultraviolet completion.
Within this perspective, the Peccei--Quinn symmetry should be viewed as an
accidental low-energy symmetry whose breaking is governed primarily by soft
supersymmetry-breaking dynamics.

Looking ahead, the framework presented here opens several avenues for further
investigation. On the experimental side, future searches for heavy axion-like
particles, displaced decays, and rare processes may probe the parameter regions
identified in this work. On the theoretical side, a more complete treatment of the
early-Universe dynamics of the axion supermultiplet, including saxion and axino
production and decay, would enable a quantitative assessment of the long-lived
axion regime. Embedding the PQ sector into a fully specified supersymmetric model
with an explicit ultraviolet completion would further sharpen predictions for
axion couplings and clarify the relation between supersymmetry breaking and axion
physics.

The analysis presented here is intentionally agnostic about the microscopic origin
of a small effective QCD vacuum angle. The axion-like particle considered in this
framework does not provide a dynamical solution to the strong CP problem in the
traditional Peccei--Quinn sense. Instead, the axion mass is treated as being
dominated by soft supersymmetry-breaking effects, independently of whatever
mechanism suppresses the strong CP phase in the ultraviolet. In this respect, the
framework should be interpreted as a consistent and predictive realization of
heavy axion-like particle physics rather than as a proposal for solving the strong
CP problem.

A global summary of the phenomenological implications is provided in
Fig.~\ref{fig:likelihood_exclusion}, where representative likelihood contours are
shown together with laboratory and cosmological constraints. The remaining
parameter regions define concrete benchmark scenarios that can be explored by
future intensity-frontier experiments and collider facilities.

In summary, we have demonstrated that soft supersymmetry breaking can serve as the
sole origin of axion masses within a controlled effective field theory framework,
leading to a predictive and phenomenologically rich axion-like particle scenario.
The close interplay between supersymmetry breaking, axion phenomenology, and
cosmology provides a distinctive setting that can be tested by a combination of
laboratory experiments and cosmological observations.

\appendix

\section{Minimization of the PQ scalar potential}
\label{app:minimization}

In this appendix, we present the details of the minimization of the scalar potential
of the Peccei--Quinn (PQ) sector and demonstrate the existence of a stable vacuum
that spontaneously breaks the PQ symmetry.

The scalar potential receives contributions from supersymmetric $F$-terms and soft
supersymmetry-breaking terms,
\begin{equation}
V(S) = |\kappa S^2|^2
+ m_S^2 |S|^2
+ \left(
\frac{1}{3} A_\kappa \kappa S^3
+ \frac{1}{2} B_S S^2
+ \text{h.c.}
\right).
\end{equation}

We parametrize the complex scalar field as
\begin{equation}
S = \frac{1}{\sqrt{2}}\, v_s e^{i\theta}.
\end{equation}

The stationary condition with respect to $v_s$ yields
\begin{equation}
\frac{\partial V}{\partial v_s} =
\kappa^2 v_s^3
+ m_S^2 v_s
+ A_\kappa \kappa v_s^2
+ B_S v_s = 0 .
\end{equation}

A non-trivial solution exists for $m_S^2<0$ and is approximately given by
\begin{equation}
v_s \simeq \frac{1}{\kappa}
\left(
- A_\kappa \pm \sqrt{A_\kappa^2 - 4 m_S^2}
\right).
\end{equation}

Stability of the vacuum is ensured by the condition
\begin{equation}
\frac{\partial^2 V}{\partial v_s^2}
= 3 \kappa^2 v_s^2 + m_S^2 + 2 A_\kappa \kappa v_s > 0 .
\end{equation}

\section{Axion supermultiplet mass matrices}
\label{app:masses}

Expanding around the vacuum, the PQ field is written as
\begin{equation}
S = \frac{1}{\sqrt{2}} (v_s + \sigma + i a).
\end{equation}

The scalar mass terms can be written as
\begin{equation}
\mathcal{L}_{\text{mass}}^{\text{scalar}}
= \frac{1}{2}
\begin{pmatrix}
a & \sigma
\end{pmatrix}
\begin{pmatrix}
4 B_S & 0 \\
0 & 2 \kappa^2 v_s^2 + m_S^2
\end{pmatrix}
\begin{pmatrix}
a \\ \sigma
\end{pmatrix}.
\end{equation}

This yields
\begin{equation}
m_a^2 = 4 B_S,
\qquad
m_\sigma^2 = 2 \kappa^2 v_s^2 + m_S^2 .
\end{equation}

The axino mass follows from the superpotential interaction,
\begin{equation}
\mathcal{L} \supset -\kappa S \tilde a \tilde a
\quad \Rightarrow \quad
m_{\tilde a} = \sqrt{2}\,\kappa v_s .
\end{equation}


\section{Axion vacuum structure in the presence of multiple mass sources}
\label{app:CP}

In this appendix we discuss the structure of the axion potential in the presence of
multiple contributions to the axion mass. This analysis is intended to clarify the
interplay between soft supersymmetry-breaking effects and QCD-induced contributions,
without addressing the origin or suppression of the strong CP phase.

The axion potential receives contributions from soft supersymmetry-breaking terms
and from non-perturbative QCD effects,
\begin{equation}
V(a) =
m_{a,\mathrm{soft}}^2 f_a^2
\cos\!\left(\frac{a}{f_a} + \delta_{\rm soft}\right)
+
m_{a,\mathrm{QCD}}^2 f_a^2
\left[1-\cos\!\left(\frac{a}{f_a}\right)\right],
\end{equation}
where $\delta_{\rm soft}$ denotes a generic CP-violating phase associated with the
soft supersymmetry-breaking sector.

In the regime relevant for this work, the axion mass is dominated by the soft
supersymmetry-breaking contribution,
\begin{equation}
m_{a,\mathrm{soft}}^2 \gg m_{a,\mathrm{QCD}}^2 ,
\end{equation}
and the location of the axion vacuum is therefore controlled primarily by the
soft-breaking sector. The resulting vacuum expectation value of the axion field is
set by the minimization of the full potential and depends on the relative phases
and magnitudes of the contributing terms.

We emphasize that the present analysis does not rely on any alignment mechanism
between the phases of the soft supersymmetry-breaking sector and the QCD vacuum
angle. The framework explored in this work is agnostic about the microscopic origin
of a small effective strong CP phase, which may arise from ultraviolet physics or
from an independent mechanism not specified here. Accordingly, no claim is made
that the axion-like particle considered in this framework dynamically relaxes or
suppresses the strong CP angle.

\section{Radiative corrections and robustness of the spectrum}
\label{app:radiative}

Radiative corrections to the scalar potential are described by the
Coleman--Weinberg potential,
\begin{equation}
V_{\rm CW} =
\frac{1}{64\pi^2}
\sum_i (-1)^{F_i}
m_i^4(S)
\ln\!\left(\frac{m_i^2(S)}{\mu^2}\right).
\end{equation}

The induced correction to the axion mass is parametrically
\begin{equation}
\delta m_a^2
\sim \frac{1}{16\pi^2}
\frac{m_{3/2}^2}{f_a^2}
\ll m_a^2 ,
\end{equation}
which ensures
\begin{equation}
\frac{\delta m_a^2}{m_a^2}
\ll 1 .
\end{equation}
\section{Planck-suppressed Peccei--Quinn violation}
\label{app:gravity}

Accidental Peccei--Quinn symmetries protected by discrete gauge symmetries or
string-theoretic selection rules can suppress Planck-scale violations to high
operator dimension. Planck-suppressed operators of the form
$S^n/M_{\rm Pl}^{n-4}$ with $n \gtrsim 8$ induce axion mass contributions that remain
parametrically subdominant compared to the soft supersymmetry-breaking effects
considered in this work, provided $f_a \lesssim 10^4~\mathrm{GeV}$.

\section*{Acknowledgments}
GG acknowledges the support of UGC--RUSA, Government of India, for carrying out this work.


\begin{thebibliography}{99}

\bibitem{PQ1977a}
R.~D.~Peccei and H.~R.~Quinn,
Phys.\ Rev.\ Lett.\ \textbf{38}, 1440 (1977).

\bibitem{PQ1977b}
R.~D.~Peccei and H.~R.~Quinn,
Phys.\ Rev.\ D \textbf{16}, 1791 (1977).


\bibitem{Irastorza2018}
I.~G.~Irastorza and J.~Redondo,
Prog.\ Part.\ Nucl.\ Phys.\ \textbf{102}, 89 (2018).

\bibitem{Ringwald2014}
A.~Ringwald,
Phys.\ Dark Univ.\ \textbf{1}, 116 (2012).

\bibitem{DiLuzio2020}
L.~Di~Luzio et al.,
Phys.\ Rept.\ \textbf{870}, 1 (2020).

\bibitem{Graham2015}
P.~W.~Graham et al.,
Ann.\ Rev.\ Nucl.\ Part.\ Sci.\ \textbf{65}, 485 (2015).
\bibitem{Weinberg1978}
S.~Weinberg,
Phys.\ Rev.\ Lett.\ \textbf{40}, 223 (1978).
\bibitem{CAST2017}
V.~Anastassopoulos et al. (CAST),
Nature Phys.\ \textbf{13}, 584 (2017).

\bibitem{CAST2024}
V.~Anastassopoulos et al. (CAST),
Nature Phys.\ \textbf{20}, 141 (2024).
\bibitem{Wilczek1978}
F.~Wilczek,
Phys.\ Rev.\ Lett.\ \textbf{40}, 279 (1978).
\bibitem{IAXO2014}
E.~Armengaud et al.,
JINST \textbf{9}, T05002 (2014).

\bibitem{IAXO2021}
E.~Armengaud et al.,
JINST \textbf{16}, P02030 (2021).

\bibitem{IAXO2025}
E.~Armengaud et al.,
JHEP \textbf{03}, 056 (2025).
\bibitem{Nilles1984}
H.~P.~Nilles,
Phys.\ Rept.\ \textbf{110}, 1 (1984).

\bibitem{HaberKane1985}
H.~E.~Haber and G.~L.~Kane,
Phys.\ Rept.\ \textbf{117}, 75 (1985).

\bibitem{Kim1987}
J.~E.~Kim,
Phys.\ Rept.\ \textbf{150}, 1 (1987).

\bibitem{ChoiKim1985}
K.~Choi and J.~E.~Kim,
Phys.\ Rev.\ D \textbf{32}, 1828 (1985).

\bibitem{Bae2013}
K.~J.~Bae et al.,
JHEP \textbf{01}, 161 (2013).

\bibitem{Chun2011}
E.~J.~Chun,
Phys.\ Rev.\ D \textbf{84}, 043509 (2011).
\bibitem{Kawasaki2005}
M.~Kawasaki et al.,
Phys.\ Rev.\ D \textbf{71}, 083502 (2005).

\bibitem{Cyburt2016}
R.~H.~Cyburt et al.,
Rev.\ Mod.\ Phys.\ \textbf{88}, 015004 (2016).

\bibitem{Poulin2017}
V.~Poulin et al.,
Phys.\ Rev.\ D \textbf{96}, 083524 (2017).

\bibitem{CMB2020}
N.~Aghanim et al. (Planck),
Astron.\ Astrophys.\ \textbf{641}, A6 (2020).
\bibitem{Jaeckel2010}
J.~Jaeckel and A.~Ringwald,
Ann.\ Rev.\ Nucl.\ Part.\ Sci.\ \textbf{60}, 405 (2010).

\bibitem{Dobrich2016}
B.~D\"obrich et al.,
Phys.\ Rev.\ D \textbf{94}, 095025 (2016).

\bibitem{Bauer2017}
M.~Bauer et al.,
JHEP \textbf{09}, 152 (2017).

\bibitem{Aloni2019}
D.~Aloni et al.,
Phys.\ Rev.\ Lett.\ \textbf{123}, 071801 (2019).
\bibitem{BarrSeckel1992}
S.~M.~Barr and D.~Seckel,
Phys.\ Rev.\ D \textbf{46}, 539 (1992).

\bibitem{Holman1992}
R.~Holman et al.,
Phys.\ Lett.\ B \textbf{282}, 132 (1992).

\bibitem{Kamionkowski1992}
M.~Kamionkowski and J.~March-Russell,
Phys.\ Lett.\ B \textbf{282}, 137 (1992).

\bibitem{Banks2010}
T.~Banks and N.~Seiberg,
Phys.\ Rev.\ D \textbf{83}, 084019 (2011).





\bibitem{NA64_2020}
D.~Banerjee et al. (NA64),
Phys.\ Rev.\ Lett.\ \textbf{125}, 081801 (2020).

\bibitem{NA64_2021}
D.~Banerjee et al. (NA64),
Phys.\ Rev.\ D \textbf{103}, 072006 (2021).

\bibitem{NA64_2023}
D.~Banerjee et al. (NA64),
Phys.\ Rev.\ Lett.\ \textbf{131}, 101801 (2023).

\bibitem{SHiP2015}
S.~Alekhin et al.,
Rept.\ Prog.\ Phys.\ \textbf{79}, 124201 (2016).

\bibitem{Belle2010}
F.~T.~Avignone et al.,
Phys.\ Rev.\ D \textbf{81}, 035002 (2010).

\bibitem{BelleII2019}
E.~Kou et al. (Belle II),
PTEP \textbf{2019}, 123C01 (2019).

\bibitem{BelleII2023}
I.~Adachi et al. (Belle II),
Phys.\ Rev.\ Lett.\ \textbf{130}, 181801 (2023).

\bibitem{Raffelt2008}
G.~G.~Raffelt,
Lect.\ Notes Phys.\ \textbf{741}, 51 (2008).

\bibitem{Payez2015}
A.~Payez et al.,
JCAP \textbf{02}, 006 (2015).

\bibitem{Daylan2017}
T.~Daylan et al.,
Phys.\ Dark Univ.\ \textbf{12}, 1 (2016).




\bibitem{Georgi1986}
H.~Georgi et al.,
Phys.\ Lett.\ B \textbf{169}, 73 (1986).

\bibitem{Craig2016}
N.~Craig et al.,
JHEP \textbf{06}, 137 (2016).

\bibitem{Hook2018}
A.~Hook et al.,
Phys.\ Rev.\ Lett.\ \textbf{120}, 261801 (2018).

\bibitem{ALPglobal2021}
J.~Bonilla et al.,
JHEP \textbf{11}, 168 (2021).

\bibitem{ALPglobal2023}
M.~Bauer et al.,
JHEP \textbf{12}, 061 (2023).

\bibitem{ALPglobal2024}
A.~Caputo et al.,
Phys.\ Rev.\ D \textbf{109}, 035012 (2024).

\end{thebibliography}
\end{document}